\documentclass[preprint,prd,aps,showpacs,showkeys,nofootinbib]{revtex4}
\usepackage{graphicx}
\begin{document}
\title{Charged lepton flavor violation in extended BLMSSM}
\author{Xing-Xing Dong$^{1}$\footnote{dxx$\_$0304@163.com},
Shu-Min Zhao$^{1}$\footnote{zhaosm@hbu.edu.cn},
Hai-Bin Zhang$^{1}$\footnote{hbzhang@hbu.edu.cn},
Tai-Fu Feng$^{1}$\footnote{fengtf@hbu.edu.cn}}

\affiliation{$^1$ Department of Physics and Technology, Hebei University, Baoding 071002, China}
\begin{abstract}
Within the extended BLMSSM, the exotic Higgs superfields $(\Phi_{NL},\varphi_{NL})$ are added to make the exotic leptons heavy, and the superfields ($Y$,$Y^\prime$) are also introduced to make exotic leptons unstable. This new model is named as the EBLMSSM. We study some charged lepton flavor violating (CLFV) processes in detail in the EBLMSSM, including $l_j\rightarrow l_i \gamma$, muon conversion to electron in nuclei, the $\tau$ decays and $h^0\rightarrow l_i l_j$. Being different from BLMSSM, some particles are redefined in this new model, such as slepton, sneutrino, exotic lepton (neutrino), exotic slepton (sneutrino) and lepton neutralino. We also introduce the mass matrices of superfields $Y$ and spinor $\tilde{Y}$ in the EBLMSSM. All of these lead to new contributions to the CLFV processes.  In the suitable parameter space, we obtain the reasonable numerical results. The results of this work will encourage physicists to explore new physics beyond the SM.
\end{abstract}

\maketitle
\section{Introduction}
The Higgs boson, an elementary particle, has been researched by the Large Hadron Collider (LHC) as one of the primary scientific goals. Combining the updated data of the ATLAS\cite{h0ATLAS} and CMS\cite{h0CMS} Collaborations, now its measured mass is $m_{h^0}=125.09\pm0.24 \rm {GeV}$\cite{h0LHC}, which represent that the Higgs mechanism is compellent. In the Standard Model (SM), the lepton-flavor number is conserved. However, the neutrino oscillation experiments\cite{neutrino1,neutrino2,neutrino3,neutrino4,neutrino5,neutrino6,neutrino7,neutrino8,neutrino9} have convinced that neutrinos possess tiny masses and mix with each other. So the individual lepton numbers $L_i=L_e,~L_\mu,~L_\tau$ are not exact symmetries at the electroweak scale. Furthermore, the presentation of the GIM mechanism makes the charged lepton flavor violating(CFLV) processes in the SM very tiny\cite{CLFV1,CLFV2,CLFV3}, such as $Br_{SM}(l_j\rightarrow l_i\gamma)\sim 10^{-55}$\cite{SMljlig}. Therefore, if we observe the CLFV processes in future experiments, it is an obvious evidence of new physics beyond the SM.

Studying the CLFV processes is an effective way to explore new physics beyond the SM. MEG Collaboration gives out the current experiment upper bound of the CLFV process $\mu\rightarrow e \gamma$, which is $Br(\mu\rightarrow e \gamma)< 5.7\times 10^{-13}$ at $90\%$ confidence level\cite{MEG}. $Br(\tau\rightarrow e \gamma)< 3.3\times 10^{-8}$ and $\;Br(\tau\rightarrow \mu \gamma)< 4.4\times 10^{-8}$ are also shown in Ref.\cite{PDG}. SINDRUM II collaboration has updated the sensitivity of the $\mu-e$ conversion rate in Au nuclei $CR(\mu\rightarrow e:\;_{79}^{197}Au)<7\times 10^{-13}$\cite{CRue}. The current experiment upper bounds for the $\tau$ decays $Br(\tau\rightarrow 3e)<2.7\times 10^{-8}$ and $Br(\tau\rightarrow 3\mu)<2.1\times 10^{-8}$ have been shown by Particle Data Group\cite{PDG}. Furthermore, a direct research for the 125.1 GeV Higgs boson decays including the CLFV, $h^0\rightarrow l_i l_j$, has been given out by the CMS Collaboration\cite{CMS1,CMS2} and ATLAS Collaboration\cite{ATLAS1}. We show the corresponding experiment upper bounds for processes $h^0\rightarrow l_i l_j$ in TABLE I.
Physicists do more research on the CLFV processes for $l_j\rightarrow l_i \gamma$ decays, $\mu-e$ conversion rates in nuclei, the $\tau$ decays and $h^0\rightarrow l_i l_j$ decays in models beyond the SM\cite{bSM1,bSM2,bSM3,bSM4,bSM5,bSM6}. In our previous work, we have studied $l_j\rightarrow l_i \gamma$, muon conversion to electron in nuclei, the $\tau$ decays and $h^0\rightarrow l_i l_j$ processes in the $\mu\nu$SSM\cite{SSM1,SSM2,SSM3}. We also have discussed $l_j\rightarrow l_i \gamma$ processes, the $\tau$ decays and muon conversion to electron in nuclei in the BLMSSM\cite{ljBL,CueBL}. In this work, we study the processes $l_j\rightarrow l_i \gamma$, $\mu-e$ conversion in Au nuclei, the $\tau$ decays and $h^0\rightarrow l_i l_j$ in the extended BLMSSM, which is named as the EBLMSSM\cite{EBL}.

Extending the MSSM with the introduced local gauged B and L, one obtains the so called BLMSSM\cite{BLMSSM1,BLMSSM2,BLMSSM3,BLMSSM4}. In the BLMSSM, the exotic lepton masses are obtained from the Yukawa couplings with the two Higgs doublets $H_u$ and $H_d$. The values of these exotic lepton masses are around 100 GeV, which can well anastomose the current experiment bounds. However, with the development of high energy physics experiments, we may obtain the heavier experiment lower bounds of the exotic lepton masses in the near future, which makes the BLMSSM model do not exist. Therefore, two exotic Higgs superfields, the $SU(2)_L$ singlets $\Phi_{NL}$ and $\varphi_{NL}$, are considered to be added in the BLMSSM. With the introduced superfields $\Phi_{NL}$ and $\varphi_{NL}$, the exotic leptons can turn heavy and should be unstable. This new model is named as the extended BLMSSM (EBLMSSM)\cite{EBL}. In order to make the exotic leptons unstable, we add superfields $Y$ and $Y'$ in the EBLMSSM.
\begin{table}[t]
\caption{ \label{tab1}  Present experiment limits for 125 GeV Higgs decays $h^0\rightarrow l_i l_j$ with the CLFV.}
\footnotesize
\begin{tabular*}{125mm}{@{\extracolsep{\fill}}cccc}
\toprule  CLFV process&Present limit(CMS)&Present limit(ATLAS)&confidence level (CL) \\
\hline
$h^0\rightarrow e \mu$ \hphantom{00} & \hphantom{0}$< 0.035\%$\cite{CMS1}& \hphantom{0}$ --$ & \hphantom{0}$95\%$ \\
$h^0\rightarrow e \tau$ \hphantom{00} & \hphantom{0}$<0.61\%$\cite{CMS2}& \hphantom{0}$<1.04\%$\cite{ATLAS1} & \hphantom{0}$95\%$ \\
$h^0\rightarrow \mu\tau$ \hphantom{00} & \hphantom{0}$<0.25\%$\cite{CMS2}& \hphantom{0}$<1.43\%$\cite{ATLAS1} &\hphantom{0} $95\%$ \\
\hline
\end{tabular*}%
\end{table}

In the BLMSSM, the dark matter (DM) candidates include the lightest mass eigenstate of $X, X^\prime$ mixing and a four-component spinor $\tilde{X}$ composed by the superpartners of $X,X^\prime$. In the EBLMSSM, the DM candidates not only include above terms presented in the BLMSSM, but also contain new terms due to the new introduced superfields of $Y,Y'$. So the lighter mass eigenstates of $Y,Y^\prime$ mixing and spinor $\tilde{Y}$ are DM candidates\cite{EBL,DM}. In section 4.2 of our previous work\cite{EBL}, we suppose the lightest mass eigenstate of $Y, Y^\prime$ mixing as a DM candidate, and calculate the relic density $\Omega_Dh^2$. In the reasonable parameter space, $\Omega_Dh^2$ of $Y_1$ can match the experiment results well.

The Higgs boson $h^0$ is produced chiefly from the gluon fusion $(gg\rightarrow h^0)$ at the LHC.
The leading order (LO) contributions originate from the one-loop diagrams. In the BLMSSM, we have studied the $ h^0\rightarrow gg$ process in our previous work\cite{soft1}, and the virtual top quark loops play the dominate roles. The EBLMSSM results for $ h^0\rightarrow gg$ are same as those in BLMSSM, which have been discussed in Ref.\cite{EBL}. Being different from BLMSSM, the exotic leptons in EBLMSSM are more heavy and the exotic sleptons of the 4-th and 5-th generations mix together to form a $4\times4$ mass matrices. The LO contributions for $h^0\rightarrow \gamma\gamma$ originate from the one-loop diagrams. In the EBLMSSM\cite{EBL}, we have studied the decay $h^0\rightarrow \gamma\gamma$ in detail. The processes $h^0\rightarrow VV, V=(Z,W)$ also have been researched in this new model. Considering the constraints from the parameter space of these researches, we study the processes $l_j\rightarrow l_i\gamma$, muon conversion to electron in Au nuclei, $\tau$ decays and $h^0\rightarrow l_il_j$ in this work.

The outline of this paper is organized as follows. In section II, we present the ingredients of the EBLMSSM by introducing its superpotential, the general soft SUSY-breaking terms, new corrected mass matrices and couplings which are different from those in the BLMSSM. In section III, we analyze the corresponding amplitudes and the branching ratios of rare CLFV processes $l_j\rightarrow l_i \gamma$, the $\tau$ decays and $h^0\rightarrow l_i l_j$ decays. We also discuss the muon conversion to electron rates in nuclei. The numerical analysis is discussed in section IV, and the conclusions are summarized in section V. The tedious formulae are collected in Appendix.

\section{introduction of the EBLMSSM}
In the EBLMSSM, the local gauge group is
$SU(3)_{C}\otimes SU(2)_{L}\otimes U(1)_{Y}\otimes U(1)_{B}\otimes U(1)_{L}$\cite{EBL,BLMSSM1,group1,group2}. We introduce the exotic Higgs superfields $\Phi_{NL}$ and $\varphi_{NL}$
with nonzero VEVs $\upsilon_{NL}$ and $\bar{\upsilon}_{NL}$\cite{750} to make the exotic leptons heavy. Accordingly, the superfields $Y$ and $Y'$ are introduced to avoid the heavy exotic leptons stable.
In TABLE II, we show the new introduced superfields in the EBLMSSM\cite{EBL}.

The corresponding superpotential of the EBLMSSM is shown here
\begin{eqnarray}
&&{\cal W}_{{EBLMSSM}}={\cal W}_{{MSSM}}+{\cal W}_{B}+{\cal W}_{L}+{\cal W}_{X}+{\cal W}_{Y}\;,
\nonumber\\
&&{\cal W}_{L}=\lambda_{L}\hat{L}_{4}\hat{L}_{5}^c\hat{\varphi}_{NL}+\lambda_{E}\hat{E}_{4}^c\hat{E}_{5}
\hat{\Phi}_{NL}+\lambda_{NL}\hat{N}_{4}^c\hat{N}_{5}\hat{\Phi}_{NL}
+\mu_{NL}\hat{\Phi}_{NL}\hat{\varphi}_{NL}\nonumber\\&&\hspace{1.2cm}+Y_{{e_4}}\hat{L}_{4}\hat{H}_{d}\hat{E}_{4}^c+Y_{{\nu_4}}\hat{L}_{4}\hat{H}_{u}\hat{N}_{4}^c
+Y_{{e_5}}\hat{L}_{5}^c\hat{H}_{u}\hat{E}_{5}+Y_{{\nu_5}}\hat{L}_{5}^c\hat{H}_{d}\hat{N}_{5}
\nonumber\\
&&\hspace{1.2cm}
+Y_{\nu}\hat{L}\hat{H}_{u}\hat{N}^c+\lambda_{{N^c}}\hat{N}^c\hat{N}^c\hat{\varphi}_{L}
+\mu_{L}\hat{\Phi}_{L}\hat{\varphi}_{L}\;,
\nonumber\\&&
{\cal W}_{Y}=\lambda_4\hat{L}\hat{L}_{5}^c\hat{Y}+\lambda_5\hat{N}^c\hat{N}_{5}\hat{Y}^\prime
+\lambda_6\hat{E}^c\hat{E}_{5}\hat{Y}^\prime+\mu_{Y}\hat{Y}\hat{Y}^\prime\;.\label{super}
\end{eqnarray}
${\cal W}_{{MSSM}}$ is the superpotential of the MSSM. ${\cal W}_{B}$ and ${\cal W}_{X}$ are same as the terms in the BLMSSM\cite{soft1,soft2}. The new terms $\lambda_{L}\hat{L}_{4}\hat{L}_{5}^c\hat{\varphi}_{NL}+\lambda_{E}\hat{E}_{4}^c\hat{E}_{5}
\hat{\Phi}_{NL}+\lambda_{NL}\hat{N}_{4}^c\hat{N}_{5}\hat{\Phi}_{NL}
+\mu_{NL}\hat{\Phi}_{NL}\hat{\varphi}_{NL}$ are added to ${\cal W}_{L}$ based on the original BLMSSM. Comparing with the ${\cal W}_{X}$ in the BLMSSM, ${\cal W}_{Y}$ is introduced in the EBLMSSM, which includes the lepton-exotic lepton-$Y$ coupling and lepton-exotic slepton-$\tilde{Y}$ coupling. These new couplings can produce one-loop diagrams influencing the CLFV decays. These new couplings can also produce one-loop diagrams contributing to the lepton
electric dipole moment(EDM) and lepton magnetic dipole moment(MDM), which will be discussed in our next work. With the 4-th and 5-th generation exotic sleptons mixing together, the $h^0(Z)$-exotic slepton-exotic slepton coupling is deduced in the EBLMSSM. In the EBLMSSM, the couplings for lepton-slepton-lepton neutralino, $h^0(Z)$-slepton-slepton, $h^0(Z)$-sneutrino-sneutrino and $h^0(Z)$-exotic lepton-exotic lepton also have new contributions to CLFV processes. In the whole, the new couplings in the EBLMSSM enrich the lepton physics in a certain degree.
\begin{table}[t]
\caption{ The new introduced superfields in the EBLMSSM beyond BLMSSM}
\begin{tabular}{|c|c|c|c|c|c|}
\hline
Superfields & $SU(3)_C$ & $SU(2)_L$ & $U(1)_Y$ & $U(1)_B$ & $U(1)_L$\\
\hline
$\hat{\Phi}_{NL}$ & 1 & 1 & 0 & 0 & -3 \\
\hline
$\hat{\varphi}_{NL}$ & 1 & 1 & 0 & 0 & 3\\
\hline
$Y$ & 1 & 1 & 0 & 0 & $2+L_4$
\\ \hline
$Y'$ & 1 & 1 & 0 & 0 & $-(2+L_4)$ \\
\hline
\end{tabular}
\label{quarks}
\end{table}

In the EBLMSSM, ${\cal W}_{Y}$ are the new terms in the superpotential. In ${\cal W}_{Y}$, $\lambda_4(\lambda_6)$ is the coupling coefficient of $Y$-lepton-exotic lepton and $\tilde{Y}$-slepton-exotic slepton couplings. We consider $\lambda_4^2(\lambda_6^2)$ is a $3\times 3$ matrix and has non-zero elements relating with the CLFV. In our following numerical analysis, we assume that $(\lambda_4^2)^{IJ}=(\lambda_6^2)^{IJ}=(Lm^2)^{IJ}$, $I(J)$ represents the $I$-th ($J$-th) generation charged lepton. When $I=J$, there is no CLFV, which has no contributions to our researched decay processes. So, only the non-diagonal elements $(Lm^2)^{IJ}(I\neq J)$ influence the numerical results of the CLFV processes. Therefore, we should take into account the effects from ${\cal W}_{Y}$ in this work.

Based on the new introduced superfields $\Phi_{NL},\varphi_{NL}, Y$ and $Y'$ in the EBLMSSM, the soft breaking terms are given out
\begin{eqnarray}
&&{\cal L}_{{soft}}^{EBLMSSM}={\cal L}_{{soft}}^{BLMSSM}
-m_{{\Phi_{NL}}}^2\Phi_{NL}^*\Phi_{NL}
-m_{{\varphi_{NL}}}^2\varphi_{NL}^*\varphi_{NL}
+(A_{{LL}}\lambda_{L}\tilde{L}_{4}\tilde{L}_{5}^c\varphi_{NL}\nonumber\\&&\hspace{2.0cm}
+A_{{LE}}\lambda_{E}\tilde{e}_{4}^c\tilde{e}_{5}\Phi_{NL}
+A_{{LN}}\lambda_{NL}\tilde{\nu}_{4}^c\tilde{\nu}_{5}\Phi_{NL}
+B_{NL}\mu_{NL}\Phi_{NL}\varphi_{NL}+h.c.)\nonumber\\&&\hspace{2.0cm}+(
A_4\lambda_4\tilde{L}\tilde{L}_{5}^cY+A_5\lambda_5\tilde{N}^c\tilde{\nu}_{5}Y^\prime
+A_6\lambda_6\tilde{e}^c\tilde{e}_{5}Y^\prime+B_{Y}\mu_{Y}YY^\prime+h.c.).
\label{soft-breaking}
\end{eqnarray}
${\cal L}_{{soft}}^{BLMSSM}$ is the soft breaking terms of the BLMSSM discussed in our previous work\cite{soft1,soft2}. Here, corresponding to the $SU(2)_L$ singlets $\Phi_{NL}$ and $\varphi_{NL}$,
we obtain the nonzero VEVs $\upsilon_{NL} $ and $\bar{\upsilon}_{NL}$ respectively. Generally,  the values of these two parameters are at TeV scale. The exotic Higgs $\Phi_{NL}$ and $\varphi_{NL}$ can be written as
\begin{eqnarray}
&&\Phi_{NL}={1\over\sqrt{2}}\Big(\upsilon_{NL}+\Phi_{NL}^0+iP_{NL}^0\Big),~~~~
\varphi_{NL}={1\over\sqrt{2}}\Big(\bar{\upsilon}_{NL}+\varphi_{NL}^0+i\bar{P}_{NL}^0\Big),
\end{eqnarray}
where $\tan\beta_{NL}=\bar{\upsilon}_{NL}/\upsilon_{NL}$ and $v_{Nlt}=\sqrt{v_{NL}^2+\bar{v}_{NL}^2}$.

Comparing with the BLMSSM, the introduced superfields $\Phi_{NL}$ and $\varphi_{NL}$ in the EBLMSSM can give corrections to the mass matrices of the slepton, sneutrino, exotic lepton, exotic neutrino, exotic slepton, exotic sneutrino and lepton neutralino. However, the mass matrices of squark, exotic quark, exotic squark used in this work are same as those in the BLMSSM\cite{soft1,quark}. We deduce the adjusted mass matrices in the EBLMSSM as follows.
\subsection{The mass matrices of slepton  and sneutrino in the EBLMSSM}
In our previous work, we can easily obtain the slepton and sneutrino mass squared matrices of the BLMSSM\cite{ljBL}. Using the replacement $\overline{\upsilon}^2_L-\upsilon^2_L\rightarrow V_L^2$ (here $V_L^2=\overline{\upsilon}^2_L-\upsilon^2_L+\frac{3}{2}(\overline{\upsilon}^2_{NL}-\upsilon^2_{NL})$) for the BLMSSM results, we acquire the mass squared matrices of slepton and sneutrino in the EBLMSSM.
\subsection{The mass matrices of exotic lepton and exotic neutrino in the EBLMSSM}
The EBLMSSM exotic leptons masses are heavier than those in the BLMSSM due to the introduction of large parameters $\upsilon_{NL}$ and $\bar{\upsilon}_{NL}$. One can obtain the mass matrix of exotic lepton in the Lagrangian:
\begin{eqnarray}
&&-{\cal L}_{L'}^{mass}=\left(\begin{array}{ll}\bar{e}_{{4R}}^\prime&\bar{e}_{{5R}}^\prime\end{array}\right)
\left(\begin{array}{ll}-{1\over\sqrt{2}}\lambda_{L}\overline{\upsilon}_{NL}&{1\over\sqrt{2}}Y_{{e_5}}\upsilon_{u}\\
-{1\over\sqrt{2}}Y_{{e_4}}\upsilon_{d}&{1\over\sqrt{2}}\lambda_{E}\upsilon_{NL}
\end{array}\right)\left(\begin{array}{l}e_{{4L}}^\prime\\e_{{5L}}^\prime\end{array}\right)+h.c.
\label{ELmass}
\end{eqnarray}
Similarly, the mass matrix of the exotic neutrinos in the EBLMSSM can be given through the Lagrangian:
\begin{eqnarray}
&&-{\cal L}_{N'}^{mass}=\left(\begin{array}{ll}\bar{\nu}_{{4R}}^\prime&\bar{\nu}_{{5R}}^\prime\end{array}\right)
\left(\begin{array}{ll}{1\over\sqrt{2}}\lambda_{L}\overline{\upsilon}_{NL}&-{1\over\sqrt{2}}Y_{{\nu_5}}\upsilon_{d}\\
{1\over\sqrt{2}}Y_{{\nu_4}}\upsilon_{u}&{1\over\sqrt{2}}\lambda_{NL}\upsilon_{NL}
\end{array}\right)\left(\begin{array}{l}\nu_{{4L}}^\prime\\\nu_{{5L}}^\prime\end{array}\right)+h.c.
\label{Nmass-matrix}
\end{eqnarray}

\subsection{The mass matrices of exotic slepton and exotic sneutrino in the EBLMSSM}
In the EBLMSSM, the exotic slepton of 4-th generation and 5-th generation mix together, and its mass matrix is $4\times4$, which is different from that in the BLMSSM. Using the superpotential in Eq.(\ref{super}) and the soft breaking terms in Eq.(\ref{soft-breaking}), the mass squared matrix for exotic slepton can be obtained through Lagrangian:
\begin{eqnarray}
&&-{\cal L}_{{\tilde{E}}}^{mass}=\tilde{E}^{\dag}\cdot\mathcal{M}^2_{\tilde{E}}\cdot\tilde{E}.
\label{sleptonp}
\end{eqnarray}
With the base $\tilde{E}^{T}=(\tilde{e}_4,\tilde{e}_4^{c*},\tilde{e}_5,\tilde{e}_5^{c*})$, we show the concrete elements of exotic slepton
mass matrix $\mathcal{M}^2_{\tilde{E}}$ in the following form
\begin{eqnarray}
&&\mathcal{M}^2_{\tilde{E}}(\tilde{e}_5^{c*}\tilde{e}_5^{c})=
\lambda_L^2\frac{\bar{\upsilon}_{NL}^2}{2}+\frac{\upsilon_u^2}{2}|Y_{e_5}|^2+M^2_{\tilde{L}_5}
-\frac{g_1^2-g_2^2}{8}(\upsilon_d^2-\upsilon_u^2)-g_L^2(3+L_4)V_L^2,
\nonumber\\&&\mathcal{M}^2_{\tilde{E}}(\tilde{e}_5^{*}\tilde{e}_5)=\lambda_E^2\frac{\upsilon_{NL}^2}{2}
 +\frac{\upsilon_u^2}{2}|Y_{e_5}|^2+M^2_{\tilde{e}_5}+\frac{g_1^2}{4}(\upsilon_d^2-\upsilon_u^2)
 +g_L^2(3+L_4)V_L^2,
\nonumber\\&&\mathcal{M}^2_{\tilde{E}}(\tilde{e}_4^{*}\tilde{e}_4)=\lambda_L^2\frac{\bar{\upsilon}_{NL}^2}{2}
+\frac{g_1^2-g_2^2}{8}(\upsilon_d^2-\upsilon_u^2)+\frac{\upsilon_d^2}{2}|Y_{e_4}|^2+M^2_{\tilde{L}_4}
+g_L^2L_4V_L^2,
\nonumber\\&&\mathcal{M}^2_{\tilde{E}}(\tilde{e}_4^{c*}\tilde{e}_4^{c})=
\lambda_E^2\frac{\upsilon_{NL}^2}{2}-\frac{g_1^2}{4}(\upsilon_d^2-\upsilon_u^2)+\frac{\upsilon_d^2}{2}|Y_{e_4}|^2+M^2_{\tilde{e}_4}
-g_L^2L_4V_L^2,
\nonumber\\&&\mathcal{M}^2_{\tilde{E}}(\tilde{e}_4^{*}\tilde{e}_5)
=\upsilon_dY_{e_4}^*\lambda_E\frac{\upsilon_{NL}}{2}+\lambda_LY_{e_5}\frac{\bar{\upsilon}_{NL}v_u}{2},
~~~\mathcal{M}^2_{\tilde{E}}(\tilde{e}_5\tilde{e}_5^{c})=\mu^*\frac{\upsilon_d}{\sqrt{2}}Y_{e_5}+A_{e_5}Y_{e_5}\frac{\upsilon_u}{\sqrt{2}},
\nonumber\\&&\mathcal{M}^2_{\tilde{E}}(\tilde{e}_4^{c}\tilde{e}_5)=\mu_{NL}^*\lambda_E
\frac{\bar{\upsilon}_{NL}}{\sqrt{2}}-A_{LE}\lambda_E\frac{\upsilon_{NL}}{\sqrt{2}},
~~\mathcal{M}^2_{\tilde{E}}(\tilde{e}_4\tilde{e}_5^{c})=-\mu_{NL}^*\frac{\upsilon_{NL}}{\sqrt{2}}\lambda_L+A_{LL}\lambda_L\frac{\bar{\upsilon}_{NL}}{\sqrt{2}},
\nonumber\\&&\mathcal{M}^2_{\tilde{E}}(\tilde{e}_4\tilde{e}_4^{c})=\mu^*\frac{\upsilon_u}{\sqrt{2}}Y_{e_4}+A_{e_4}Y_{e_4}\frac{\upsilon_d}{\sqrt{2}},
~~~\mathcal{M}^2_{\tilde{E}}(\tilde{e}_5^{c}\tilde{e}_4^{c*})
=-Y_{e_5}\lambda_E\frac{\upsilon_u\upsilon_{NL}}{2}-\lambda_LY_{e_4}^*\frac{\bar{\upsilon}_{NL}v_d}{2}. \label{SE45}
\end{eqnarray}
The matrix $Z_{\tilde{E}}$ is used to rotate exotic slepton mass matrix to mass eigenstates, which is $Z^{\dag}_{\tilde{E}}\mathcal{M}^2_{\tilde{E}} Z_{\tilde{E}}=diag(m^2_{\tilde{E}^1},m^2_{\tilde{E}^2},m^2_{\tilde{E}^3},m^2_{\tilde{E}^4})$.

In the same way, the exotic sneutrino mass squared matrix is also obtained through the Lagrangian:
\begin{eqnarray}
&&-{\cal L}_{{\tilde{N}}}^{mass}=\tilde{N}^{\dag}\cdot\mathcal{M}^2_{\tilde{N}}\cdot\tilde{N},
\label{sneutrino}
\end{eqnarray}
where the corresponding elements of the matrix $\mathcal{M}^2_{\tilde{N}}$ are
\begin{eqnarray}
&&\mathcal{M}^2_{\tilde{N}}(\tilde{\nu}_5^{c*}\tilde{\nu}_5^{c})=\lambda_L^2\frac{\bar{\upsilon}_{NL}^2}{2}
-\frac{g_1^2+g_2^2}{8}(\upsilon_d^2-\upsilon_u^2) +\frac{\upsilon_d^2}{2}|Y_{\nu_5}|^2+M^2_{\tilde{L}_5}
-g_L^2(3+L_4)V_L^2,
   \nonumber\\&&\mathcal{M}^2_{\tilde{N}}(\tilde{\nu}_4^{*}\tilde{\nu}_4)
   =\lambda_L^2\frac{\bar{\upsilon}_{NL}^2}{2}+\frac{g_1^2+g_2^2}{8}(\upsilon_d^2-\upsilon_u^2)
    +\frac{\upsilon_u^2}{2}|Y_{\nu_4}|^2+M^2_{\tilde{L}_4}
  +g_L^2L_4V_L^2,
\nonumber\\&&\mathcal{M}^2_{\tilde{N}}(\tilde{\nu}_5^{*}\tilde{\nu}_5)=
\lambda_{NL}^2\frac{\upsilon_{NL}^2}{2}+g_L^2(3+L_4)V_L^2
   +\frac{\upsilon_d^2}{2}|Y_{\nu_5}|^2+M^2_{\tilde{\nu}_5},
\nonumber\\&&\mathcal{M}^2_{\tilde{N}}(\tilde{\nu}_4^{c*}\tilde{\nu}_4^{c})=
\lambda_{NL}^2\frac{\upsilon_{NL}^2}{2}-g_L^2L_4V_L^2
   +\frac{\upsilon_u^2}{2}|Y_{\nu_4}|^2+M^2_{\tilde{\nu}_4},
\nonumber\\&&\mathcal{M}^2_{\tilde{N}}(\tilde{\nu}_5^{c}\tilde{\nu}_4^{c*})=
\lambda_{NL}Y_{\nu_5}\frac{\upsilon_{NL}\upsilon_d}{2}-\lambda_LY_{\nu_4}^*\frac{\bar{\upsilon}_{NL}\upsilon_u}{2},
~~~\mathcal{M}^2_{\tilde{N}}(\tilde{\nu}_5\tilde{\nu}_5^{c})= \mu^*\frac{\upsilon_u}{\sqrt{2}}Y_{\nu_5}+A_{\nu_5}Y_{\nu_5}\frac{\upsilon_d}{\sqrt{2}},
\nonumber\\&&
\mathcal{M}^2_{\tilde{N}}(\tilde{\nu}_4^{c}\tilde{\nu}_5)
=\mu_{NL}^*\lambda_{NL}\frac{\bar{\upsilon}_{NL}}{\sqrt{2}}-A_{LN}\lambda_N\frac{\upsilon_{NL}}{\sqrt{2}},
~~~\mathcal{M}^2_{\tilde{N}}(\tilde{\nu}_4\tilde{\nu}_5^{c})=\mu_{NL}^*\frac{\upsilon_{NL}}{\sqrt{2}}
\lambda_L-A_{LL}\lambda_L\frac{\bar{\upsilon}_{NL}}{\sqrt{2}},
\nonumber\\&&\mathcal{M}^2_{\tilde{N}}(\tilde{\nu}_4^{*}\tilde{\nu}_5)
=\lambda_LY_{\nu_5}\frac{\bar{\upsilon}_{NL}\upsilon_d}{2}-\frac{\upsilon_u\upsilon_{NL}}{2}\lambda_{NL}Y_{\nu_4}^*,~~~
\mathcal{M}^2_{\tilde{N}}(\tilde{\nu}_4\tilde{\nu}_4^{c})=\mu^*\frac{\upsilon_d}{\sqrt{2}}Y_{\nu_4}+A_{\nu_4}Y_{\nu_4}\frac{\upsilon_u}{\sqrt{2}}.
\end{eqnarray}
In the base $(\tilde{\nu}_4,\tilde{\nu}_4^{c*},\tilde{\nu}_5,\tilde{\nu}_5^{c*})$, we can diagonalize the mass squared matrix $\mathcal{M}^2_{\tilde{N}}$ by $Z_{\tilde{N}}$.
\subsection{The lepton neutralino mass matrix in the EBLMSSM}
In the EBLMSSM, $\lambda_L$, the superpartner
of the new lepton type gauge boson $Z^\mu_L$, mixes with $(\psi_{\Phi_L},\psi_{\varphi_L},\psi_{\Phi_{NL}},\psi_{\varphi_{NL}})$ (the SUSY
superpartners of the superfields ($\Phi_{L},\varphi_{L},\Phi_{NL},\varphi_{NL}$)). So the lepton neutralino mass matrix is obtained in the base $(i\lambda_L,\psi_{\Phi_L},\psi_{\varphi_L},\psi_{\Phi_{NL}},\psi_{\varphi_{NL}})$,
\begin{equation}
\mathcal{M}_L=\left(     \begin{array}{ccccc}
  2M_L &2\upsilon_Lg_L &-2\bar{\upsilon}_Lg_L&3\upsilon_{NL}g_L &-3\bar{\upsilon}_{NL}g_L\\
   2\upsilon_Lg_L & 0 &-\mu_L& 0 & 0\\
   -2\bar{\upsilon}_Lg_L&-\mu_L &0& 0 & 0\\
   3\upsilon_{NL}g_L & 0 & 0 & 0 & -\mu_{NL}\\
   -3\bar{\upsilon}_{NL}g_L& 0&0&-\mu_{NL}&0
    \end{array}\right).
\end{equation}
The mass matrix $\mathcal{M}_L$ can be diagonalized by the rotation matrix $Z_{NL}$. Then, we can have
\begin{eqnarray}
&&i\lambda_L=Z_{NL}^{1i}K_{L_i}^0
,~~~\psi_{\Phi_L}=Z_{NL}^{2i}K_{L_i}^0
,~~~\psi_{\varphi_L}=Z_{NL}^{3i}K_{L_i}^0,
\nonumber\\&&\psi_{\Phi_{NL}}=Z_{NL}^{4i}K_{L_i}^0
,~~~~~\psi_{\varphi_{NL}}=Z_{NL}^{5i}K_{L_i}^0.
\end{eqnarray}
Here, $X^0_{L_i}=(K_{L_i}^0,\bar{K}_{L_i}^0)^T$ represent the mass egeinstates of the lepton neutralino.
\subsection{The superfields $Y$ in the EBLMSSM}
The scalar superfields $Y$ and $Y'$ mix. Adopting the unitary transformation,
  \begin{eqnarray}
 \left(     \begin{array}{c}
  Y_{1} \\  Y_{2}\\
    \end{array}\right) =Z_{Y}^{\dag}\left( \begin{array}{c}
  Y \\  Y'^*\\
    \end{array}\right),
       \end{eqnarray}
the mass squared matrix for the superfield $Y$ is deduced. With $S_{Y}=g_{L}^2(2+L_{4})V_L^2$,
the concrete form for the $Y$ mass squared matrix is shown here
 \begin{eqnarray}
{\cal M}_Y^2=\left(     \begin{array}{cc}
  |\mu_{Y}|^2+S_{Y} &-\mu_{Y}B_{Y} \\
    -\mu^*_{Y}B^*_{Y} & |\mu_{Y}|^2-S_{Y}\\
    \end{array}\right).
   \end{eqnarray}
The matrix $Z_Y$ is used to diagonalize the matrix to the mass eigenstates:
  \begin{eqnarray}
Z^{\dag}_{Y}\left(     \begin{array}{cc}
  |\mu_{Y}|^2+S_{Y} &-\mu_{Y}B_{Y} \\
    -\mu^*_{Y}B^*_{Y} & |\mu_{Y}|^2-S_{Y}\\
    \end{array}\right)  Z_{Y}=\left(     \begin{array}{cc}
 m_{{Y_1}}^2 &0 \\
    0 & m_{{Y_2}}^2\\
    \end{array}\right).\label{YY'}
   \end{eqnarray}
We suppose $m_{{Y_1}}^2<m_{{Y_2}}^2$. The superpartners of $Y$ and $Y'$ form a four-component Dirac spinor $\tilde{Y}$, and the mass term for superfield $\tilde{Y}$ in the Lagrangian is given out
\begin{eqnarray}
  &&-\mathcal{L}^{mass}_{\tilde{Y}}=\mu_Y\bar{\tilde{Y}}\tilde{Y}
  ,~~~~~~~~~~~~~~~~\tilde{Y} =\left( \begin{array}{c}
  \psi_{Y'} \\  \bar{\psi}_{Y}\\
    \end{array}\right).
\end{eqnarray}

In the EBLMSSM, the 4-th and 5-th generation exotic sleptons mix together.
So the exotic slepton couplings in this new model are different from those
in the BLMSSM. We deduce the $h^0$-exotic slepton-exotic slepton ($h^0-\tilde{E}-\tilde{E}$) coupling as follows
\begin{eqnarray}
&&\mathcal{L}_{h^0\tilde{E}\tilde{E}}=\sum_{i,j=1}^4\tilde{E}^{i*}\tilde{E}^{j}h^0\Big[\Big(e^2\upsilon\sin\beta\frac{1-4s_W^2}{4s_W^2c_W^2}(Z_{\tilde{E}}^{4i*}Z_{\tilde{E}}^{4j}
-Z_{\tilde{E}}^{1i*}Z_{\tilde{E}}^{1j})
    -\frac{\mu^*}{\sqrt{2}}Y_{e_4}Z_{\tilde{E}}^{2i*}Z_{\tilde{E}}^{1j}\nonumber\\&&
    -\upsilon\sin\beta|Y_{e_5}|^2\delta_{ij}-\frac{A_{E_5}}{\sqrt{2}}Z_{\tilde{E}}^{4i*}Z_{\tilde{E}}^{3j}
    +\frac{1}{2}\lambda_LY_{e_5}Z_{\tilde{E}}^{3j}Z_{\tilde{E}}^{3i*}\bar{\upsilon}_{NL}
-\frac{1}{2}Y_{e_5}^*Z_{\tilde{E}}^{4j}\lambda_EZ_{\tilde{E}}^{2i*}\upsilon_{NL}\Big)\cos\alpha\nonumber\\&&-
    \Big(e^2\upsilon\cos\beta\frac{1-4s_W^2}{4s_W^2c_W^2}(Z_{\tilde{E}}^{1i*}Z_{\tilde{E}}^{1j}-Z_{\tilde{E}}^{4i*}Z_{\tilde{E}}^{4j})
  -\upsilon\cos\beta|Y_{e_4}|^2\delta_{ij}-\frac{A_{E_4}}{\sqrt{2}}Z_{\tilde{E}}^{2i*}Z_{\tilde{E}}^{1j}\nonumber\\&&
    -\frac{\mu^*}{\sqrt{2}}Y_{e_5}Z_{\tilde{E}}^{4i*}Z_{\tilde{E}}^{3j}-\frac{1}{2}Y_{e_4}^*Z_{\tilde{E}}^{2j}\lambda_LZ_{\tilde{E}}^{4i*}\bar{\upsilon}_{NL}
+\frac{1}{2}Z_{\tilde{E}}^{1i*}Y_{e_4}^*\lambda_EZ_{\tilde{E}}^{3j}\upsilon_{NL}
    \Big)\sin\alpha\Big].\label{hEECP}
\end{eqnarray}
We also deduce the $Z$-exotic slepton-exotic slepton ($Z-\tilde{E}-\tilde{E}$) coupling in the EBLMSSM, which is given out as
\begin{eqnarray}
&&\mathcal{L}_{Z\tilde{E}\tilde{E}}=\sum_{i,j=1}^4\tilde{E}^{i*}\tilde{E}^{j}Z\frac{e}{2s_Wc_W}\Big[
\Big(Z_{\tilde{E}}^{1i*}Z_{\tilde{E}}^{1j}+Z_{\tilde{E}}^{4i*}Z_{\tilde{E}}^{4j}\Big)
-2s_W^2\delta_{ij}\Big].\label{ZEECP}
\end{eqnarray}

As the new introduced superfield in the EBLMSSM, $Y$ leads to new couplings. The lepton-exotic lepton-$Y$ coupling used in this work is shown here
\begin{eqnarray}
&&\mathcal{L}_{lL'Y}=\sum_{i,j=1}^2\bar{l}^I\Big(\lambda_4W_L^{1i}Z_Y^{1j*}P_R -\lambda_6U_L^{2i}Z_Y^{2j*}P_L\Big)L'_{i+3}Y_j^*+h.c.\label{TCLY}
\end{eqnarray}

Superfield $\tilde{Y}$ is also a new term beyond the BLMSSM. We deduce the lepton-exotic slepton-$\tilde{Y}$ coupling as
\begin{eqnarray}
&&\mathcal{L}_{l\tilde{E}\tilde{Y}}=\bar{\tilde{Y}}\Big(\lambda_4Z_{\tilde{E}}^{4i*}P_L -\lambda_6Z_{\tilde{E}}^{3i*}P_R\Big)l^I\tilde{E}_i^*+h.c.
\end{eqnarray}

In the EBLMSSM, the new effects are added from the couplings of lepton-slepton-lepton neutralino, $h^0(Z)$-slepton-slepton, $h^0(Z)$-sneutrino-sneutrino, $h^0(Z)$-exotic lepton-exotic lepton and $h^0(Z)$-exotic neutrino-exotic neutrino, which are different from those in the BLMSSM. However, these couplings possess the same writing forms as those in the BLMSSM.
\section{The processes $l_j\rightarrow l_i \gamma$, muon conversion to electron in nuclei, the $\tau$ decays and $h^0\rightarrow l_i l_j$ in the EBLMSSM}
In this section, we analyze the branching ratios of CLFV processes $l_j\rightarrow l_i \gamma$, muon conversion rates to electron in Au nuclei, the branching ratios of rare $\tau$ decays and $h^0\rightarrow l_i l_j$ in the EBLMSSM.
\subsection{Rare decays $l_j\rightarrow l_i \gamma$}
\begin{figure}[t]
\centering
\includegraphics[width=11cm]{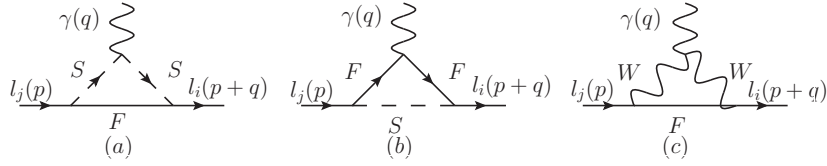}\\
\caption{The triangle type diagrams for decays $l_j\rightarrow l_i \gamma$.} \label{fig1}
\end{figure}

Generally, the corresponding effective amplitude for processes $l_j\rightarrow l_i\gamma$ can be written as\cite{efflj}
\begin{eqnarray}
&&{\cal M}=e\epsilon^{\mu}{\bar u}_i(p+q)[q^2\gamma_{\mu}(C_1^LP_L+C_1^RP_R)+m_{l_j}i\sigma_{\mu\nu}q^{\nu}(C_2^LP_L+C_2^RP_R)]u_j(p),
\nonumber\\&&C_{\alpha}^{L,R}=C_{\alpha}^{L,R}(n)+C_{\alpha}^{L,R}(c)+C_{\alpha}^{L,R}(W),\alpha=1,2,
\end{eqnarray}
where $p$ ($q$) represents the injecting lepton (photon) momentum. $m_{l_j}$ is the j-th generation lepton mass. $\epsilon$ is the photon polarization vector and $u_i(p)$ ($v_i(p)$) is the lepton (antilepton) wave function. In FIG.\ref{fig1}, we show the relevant Feynman diagrams corresponding to above amplitude. The Wilson coefficients $C_{\alpha}^{L,R}(\alpha=1,2)$ are discussed as follows.

$C_{\alpha}^{L,R}(n)(\alpha=1,2)$, the virtual neutral fermion contributions corresponding to FIG.\ref{fig1}(a), are deduced in the following form,
\begin{eqnarray}
&&C_1^L(n)=\sum_{F=\chi^0/\chi_L^0,\nu,\tilde{Y}}\sum_{S=\tilde{L},H^{\pm},\tilde{E}}\frac{1}{6m_{\Lambda}^2}
H_R^{SF\bar{l}_i}H_L^{S^*l_j\bar{F}}I_4(x_F,x_S),
\nonumber\\&&C_2^L(n)=\sum_{F=\chi^0/\chi_L^0,\nu,\tilde{Y}}\sum_{S=\tilde{L},H^{\pm},\tilde{E}}\frac{m_F}{m_{l_j}m_{\Lambda}^2}
H_L^{SF\bar{l}_i}H_L^{S^*l_j\bar{F}}[I_2(x_F,x_S)-I_3(x_F,x_S)],
\nonumber\\&&C_{\alpha}^R(n)=C_{\alpha}^L(n)|_{L{\leftrightarrow}R},\alpha=1,2,
\end{eqnarray}
where $x_i=m_i^2/m_{\Lambda}^2$, $m_i$ is the corresponding particle mass and $m_{\Lambda}$ is the new physics energy scale. $H_{L,R}^{SF\bar{l}_i}$ represent the left (right)-hand part of the coupling vertex. The concrete expressions for one-loop functions $I_i(x_1,x_2)(i=1,2,3,4,5)$ are collected in Appendix.

Then, we discuss the virtual charged fermion contributions $C_{\alpha}^{L,R}(c)(\alpha=1,2)$ corresponding to FIG.\ref{fig1}(b)
\begin{eqnarray}
&&C_1^L(c)=\sum_{F=\chi^{\pm},L'}\sum_{S=\tilde{\nu},Y}\frac{1}{6m_{\Lambda}^2}
H_R^{SF\bar{l}_i}H_L^{S^*l_j\bar{F}}[3I_2(x_S,x_F)-I_4(x_S,x_F)],
\nonumber\\&&C_2^L(c)=\sum_{F=\chi^{\pm},L'}\sum_{S=\tilde{\nu},Y}\frac{m_F}{m_{l_j}m_{\Lambda}^2}
H_L^{SF\bar{l}_i}H_L^{S^*l_j\bar{F}}I_2(x_S,x_F),
\nonumber\\&&C_{\alpha}^R(c)=C_{\alpha}^L(c)|_{L{\leftrightarrow}R},\alpha=1,2
\end{eqnarray}

Furthermore, the corrections from FIG.\ref{fig1}(c) are denoted by $C_{\alpha}^{L,R}(W)(\alpha=1,2)$
\begin{eqnarray}
&&C_1^L(W)=\sum_{F=\nu}\frac{1}{m_{\Lambda}^2}
H_L^{WF\bar{l}_i}H_L^{W^*l_j\bar{F}}[-2I_2(x_F,x_W)+\frac{1}{3}I_4(x_F,x_W)],
\nonumber\\&&C_2^L(W)=\sum_{F=\nu}\frac{1}{m_{\Lambda}^2}
H_L^{WF\bar{l}_i}H_L^{W^*l_j\bar{F}}\frac{m_{l_i}}{m_{l_j}}[I_2(x_F,x_W)+I_4(x_F,x_W)],
\nonumber\\&&C_1^R(W)=0,
\nonumber\\&&C_2^R(W)=\sum_{F=\nu}\frac{1}{m_{\Lambda}^2}
H_L^{WF\bar{l}_i}H_L^{W^*l_j\bar{F}}[I_2(x_F,x_W)+I_4(x_F,x_W)].
\end{eqnarray}
However, the contributions from $W$-$W$-neutrino diagram can be ignored due to the tiny neutrino mass.

We deduce the decay widths for processes $l_j\rightarrow l_i \gamma$
\begin{eqnarray}
&&\Gamma\left(l_j\rightarrow l_i\gamma\right)=\frac{e^2}{16\pi}m_{l_j}^{5}\left(|C_2^L|^2+|C_2^R|^2\right).
\end{eqnarray}
Then, the concrete branching ratios of $l_j\rightarrow l_i \gamma$ can be expressed as\cite{efflj}
\begin{eqnarray}
&&Br\left(l_j\rightarrow l_i\gamma\right)=\Gamma\left(l_j\rightarrow l_i\gamma\right)/\Gamma_{l_j}.
\end{eqnarray}
Here, $\Gamma_{l_j}$ represent the total decay widths of the charged leptons $l_j$. We take $\Gamma_{\mu}\simeq2.996\times10^{-19}$ GeV and $\Gamma_{\tau}\simeq2.265\times10^{-12}$ GeV\cite{PDG} in our latter numerical calculations.
\subsection{$\mu-e$ conversion in Au nuclei within the EBLMSSM}
\begin{figure}[t]
\centering
\includegraphics[width=3cm]{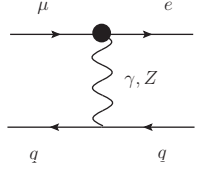}\\
\caption{The penguin type diagrams for the $\mu-e$ conversion processes at the quark level.} \label{figCp}
\end{figure}

\begin{figure}[t]
\centering
\includegraphics[width=11cm]{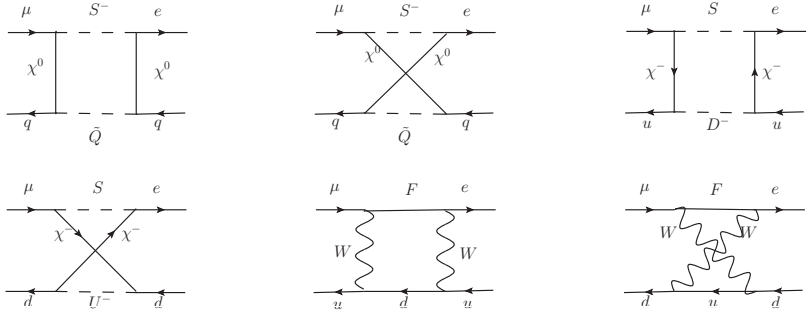}\\
\caption{The box type diagrams for the $\mu-e$ conversion processes at the quark level.} \label{figCb}
\end{figure}

In this section, we just give out the figures for $\mu-e$ conversion in nuclei at the quark level within the EBLMSSM, which are shown in FIG.\ref{figCp} and FIG.\ref{figCb}. In the BLMSSM, the theoretical results for muon conversion to electron rates in nuclei are discussed specifically in our previous work\cite{SSM3,CueBL}. We find that Au nuclei currently give the most stringent bound on conversion rates, so we only study the $\mu-e$ conversion rates in Au nuclei in this work. The new corrected particles in the EBLMSSM play important roles to this $\mu-e$ conversion processes. Considering the constraints from $\mu\rightarrow e\gamma$ within EBLMSSM, we study $\mu-e$ conversion in Au nuclei, and the corresponding numerical results will be discussed in subsection B of section IV.
\subsection{Rare $\tau$ decays within the EBLMSSM}
\begin{figure}[t]
\centering
\includegraphics[width=3cm]{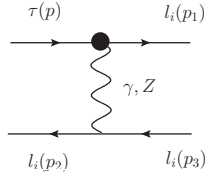}\\
\caption{The penguin type diagrams for the $\tau$ decays.} \label{tyo3ep1}
\end{figure}

\begin{figure}[t]
\centering
\includegraphics[width=11cm]{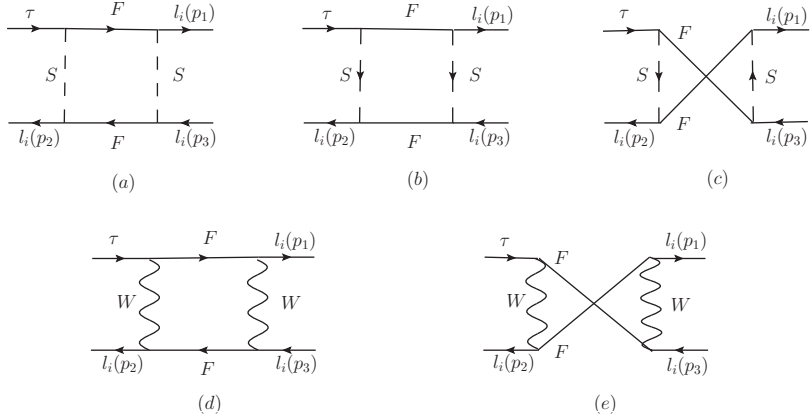}\\
\caption{The box type diagrams for the $\tau$ decays.} \label{tyo3eb1}
\end{figure}

In this section, we discuss the rare $\tau$ decays, which are $\tau\rightarrow 3l_i$ and $l_i$ represents particle $e$ or $\mu$. We give out both the penguin type diagrams and box type diagrams in FIG.\ref{tyo3ep1} and FIG.\ref{tyo3eb1}. The theoretical results for the $\tau$ decays are discussed specifically in our previous work\cite{ljBL}. In the EBLMSSM, the numerical results of $\tau$ decays can be influenced by the new corrected particles, such as exotic lepton (slepton), slepton (sneutrino), lepton neutralino, $Y$ and $\tilde{Y}$. In our latter work, we will analyze this $\tau$ decays in detail.
\subsection{Rare decay $h^0\rightarrow l_i l_j$}
\begin{figure}[t]
\centering
\includegraphics[width=14cm]{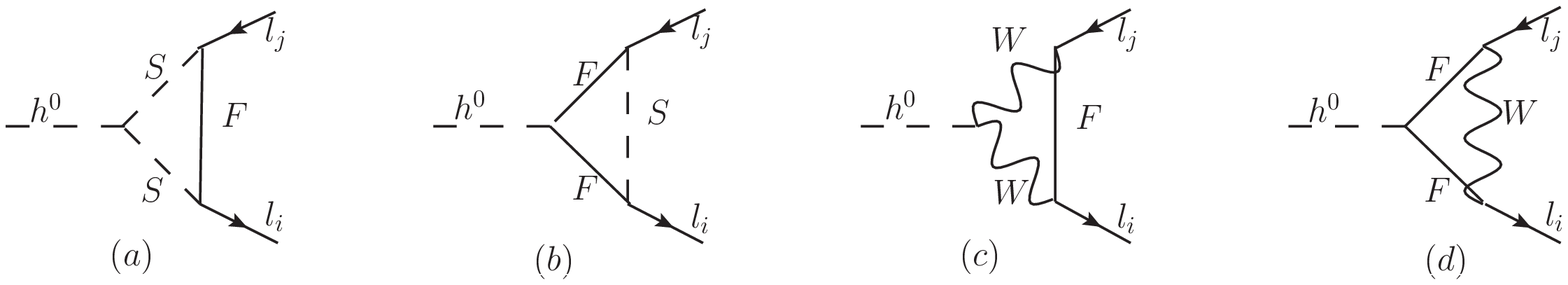}\\
\caption{The triangle type diagrams for decays $h^0\rightarrow \bar{l}_i l_j$.} \label{fig2}
\end{figure}
\begin{figure}[t]
\centering
\includegraphics[width=14cm]{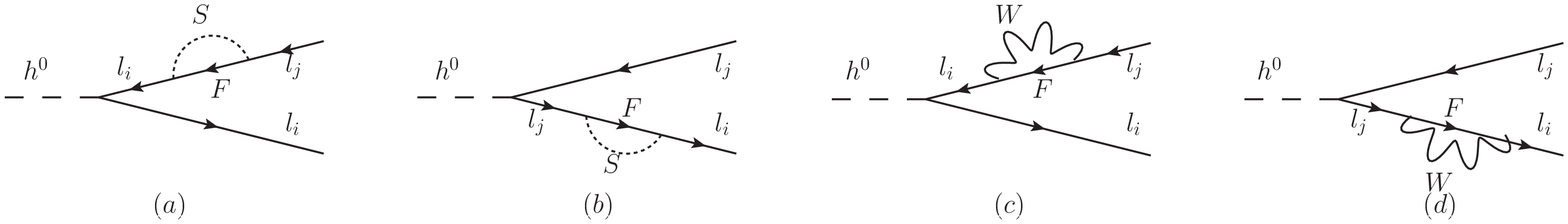}\\
\caption{The self-energy type diagrams for decays $h^0\rightarrow \bar{l}_i l_j$.} \label{fig3}
\end{figure}
The corresponding effective amplitude for $h^0\rightarrow \bar{l}_i l_j$ can be summarized as
\begin{eqnarray}
&&{\cal A}={\bar u}_i(q)(N_LP_L+N_RP_R)v_j(p),
\nonumber\\&&N_{L,R}=N_{L,R}(S_1)+N_{L,R}(S_2)+N_{L,R}(W)
\nonumber\\&&\hspace{1cm}+A_{L,R}(S_1)+A_{L,R}(S_2)+A_{L,R}(W_1)+A_{L,R}(W_2).
\end{eqnarray}
Here $N_{L,R}(S_1)$ are the coupling coefficients corresponding to triangle diagrams in FIG.\ref{fig2}(a), $N_{L,R}(S_2)$ denote the contributions from FIG.\ref{fig2}(b). The effects from FIG.\ref{fig2}(c) and FIG.\ref{fig2}(d) can be shown by $N_{L,R}(W)$. $A_{L,R}(S_1)$ and $A_{L,R}(S_2)$ represent the contributions from self-energy diagrams FIG.\ref{fig3}(a) and FIG.\ref{fig2}(b) respectively. The effects from FIG.\ref{fig3}(c) and FIG.\ref{fig3}(d) can be summarized by $A_{L,R}(W_1)$ and $A_{L,R}(W_2)$ respectively. We give out the concrete expressions for these contributions as follows.

The contributions from triangle diagrams in FIG.\ref{fig2}:
\begin{eqnarray}
&&N_L(S_1)=\sum_{F=\chi^{\pm}\chi^0/\chi_L^0,\nu,\tilde{Y}}\sum_{S=\tilde{\nu},\tilde{L},H^{\pm},\tilde{E}}
\frac{m_F}{m_{\Lambda}^2}H_L^{S_2F\bar{l}_i}H^{h^0S_1S_2^*}H_L^{S_1^*l_j\bar{F}}G_1(x_F,x_{S_1},x_{S_2}),
\nonumber\\&&N_R(S_1)=N_L(S_1)|_{L{\leftrightarrow}R},
\end{eqnarray}
\begin{eqnarray}
&&N_L(S_2)=\sum_{F=\chi^{\pm},\chi^0,\nu,L'}\sum_{S=\tilde{\nu},\tilde{L},H^{\pm},Y}
\big[H_L^{SF_2\bar{l}_i}H_R^{h^0F_1\bar{F}_2}H_L^{S^*l_j\bar{F}_1}G_2(x_S,x_{F_1},x_{F_2})
\nonumber\\&&\hspace{1.5cm}+\frac{m_{F_1}m_{F_2}}{m_{\Lambda}^2}
H_L^{SF_2\bar{l}_i}H_L^{h^0F_1\bar{F}_2}H_L^{S^*l_j\bar{F}_1}G_1(x_S,x_{F_1},x_{F_2})\big],
\nonumber\\&&N_R(S_2)=N_L(S_2)|_{L{\leftrightarrow}R},
\end{eqnarray}
\begin{eqnarray}
&&N_L(W)=-\sum_{F=\nu}\frac{m_{l_i}}{m_{\Lambda}^2}H_L^{WF\bar{l}_i}H_L^{\bar{F}l_jW^*}H^{h^0WW^*}I_2(x_F,x_W)
\nonumber\\&&+\hspace{-0.2cm}\sum_{F_1,F_2=\nu}\hspace{-0.2cm}\sqrt{x_{l_i}x_{F_2}}H_L^{WF_2\bar{l}_i}H_L^{h^0F_1\bar{F}_2}H_L^{\bar{F}_1l_jW^*}
\hspace{-0.1cm}[G_1(x_W,x_{F_1},x_{F_2})\hspace{-0.1cm}+x_{F_2}\frac{d}{dx_{F_2}}G_1(x_W,x_{F_1},x_{F_2})],
\nonumber\\&&N_R(W)=-\sum_{F=\nu}\frac{m_{l_j}}{m_{\Lambda}^2}H_L^{WF\bar{l}_i}H_L^{\bar{F}l_jW^*}H^{h^0WW^*}I_2(x_F,x_W)
\nonumber\\&&+\hspace{-0.2cm}\sum_{F_1,F_2=\nu}\hspace{-0.2cm}\sqrt{x_{l_j}x_{F_2}}H_L^{WF_2\bar{l}_i}H_L^{h^0F_1\bar{F}_2}H_L^{\bar{F}_1l_jW^*}
\hspace{-0.1cm}[G_1(x_W,x_{F_1},x_{F_2})\hspace{-0.1cm}+x_{F_1}\frac{d}{dx_{F_1}}G_1(x_W,x_{F_1},x_{F_2})].
\end{eqnarray}

The contributions from self-energy type diagrams correspond to FIG.\ref{fig3}:
\begin{eqnarray}
&&A_L(S_1)=\sum_{F=\chi^0/\chi_L^0,\chi^{\pm},L',\tilde{Y}}\sum_{S=\tilde{L},\tilde{\nu},Y,\tilde{E}}
\frac{1}{m_{l_j}^2-m_{l_i}^2}
H^{h^0l_i\bar{l}_i}\{m_F(m_{l_j}H_R^{\bar{l}_iFS}H_R^{l_j\bar{F}S^*}
\nonumber\\&&\hspace{1.4cm}+m_{l_i}H_L^{\bar{l}_iFS}H_L^{l_j\bar{F}S^*})
[I_1(x_F,x_S)+\frac{m_{l_j}^2}{m_{\Lambda}^2}(I_2(x_F,x_S)-I_3(x_F,x_S))]
\nonumber\\&&\hspace{1.4cm}-\frac{1}{2}(m_{l_j}^2H_R^{\bar{l}_iFS}H_L^{l_j\bar{F}S^*}+m_{l_i}m_{l_j}H_L^{\bar{l}_iFS}H_R^{l_j\bar{F}S^*})
I_5(x_F,x_S)\},
\nonumber\\&&A_R(S_1)=A_L(S_1)|_{L{\leftrightarrow}R},
\end{eqnarray}
\begin{eqnarray}
&&A_L(S_2)=\sum_{F=\chi^0/\chi_L^0,\chi^{\pm},L',\tilde{Y}}\sum_{S=\tilde{L},\tilde{\nu},Y,\tilde{E}}
\frac{1}{m_{l_i}^2-m_{l_j}^2}
H^{h^0l_j\bar{l}_j}\{m_F(m_{l_i}H_R^{\bar{l}_iFS}H_R^{l_j\bar{F}S^*}
\nonumber\\&&\hspace{1.4cm}+m_{l_j}H_L^{\bar{l}_iFS}H_L^{l_j\bar{F}S^*})
[I_1(x_F,x_S)+\frac{m_{l_i}^2}{m_{\Lambda}^2}(I_2(x_F,x_S)-I_3(x_F,x_S))]
\nonumber\\&&\hspace{1.4cm}-\frac{1}{2}(m_{l_i}^2H_L^{\bar{l}_iFS}H_R^{l_j\bar{F}S^*}+m_{l_i}m_{l_j}H_R^{\bar{l}_iFS}H_L^{l_j\bar{F}S^*})
I_5(x_F,x_S)\},
\nonumber\\&&A_R(S_2)=A_L(S_2)|_{L{\leftrightarrow}R},
\end{eqnarray}
\begin{eqnarray}
&&A_L(W_1)=-\sum_{F=\nu}\frac{m_{l_j}^2}{m_{l_j}^2-m_{l_i}^2}
H^{h^0l_i\bar{l}_i}H_L^{\bar{l}_iFW}H_L^{l_j\bar{F}W^*}I_5(x_F,x_W),
\nonumber\\&&A_R(W_1)=-\sum_{F=\nu}\frac{m_{l_i}m_{l_i}}{m_{l_j}^2-m_{l_i}^2}
H^{h^0l_j\bar{l}_j}H_L^{\bar{l}_iFW}H_L^{l_j\bar{F}W^*}I_5(x_F,x_W).
\end{eqnarray}

\begin{eqnarray}
&&A_L(W_2)=-\sum_{F=\nu}\frac{m_{l_i}m_{l_j}}{m_{l_i}^2-m_{l_j}^2}
H^{h^0l_j\bar{l}_j}H_L^{\bar{l}_iFW}H_L^{l_j\bar{F}W^*}I_5(x_F,x_W),
\nonumber\\&&A_R(W_2)=-\sum_{F=\nu}\frac{m_{l_i}^2}{m_{l_i}^2-m_{l_j}^2}
H^{h^0l_j\bar{l}_j}H_L^{\bar{l}_iFW}H_L^{l_j\bar{F}W^*}I_5(x_F,x_W),
\end{eqnarray}
where, the one-loop functions $G_i(x_1,x_2,x_3)(i=1,2)$ are collected in Appendix.

The decay widths for processes $h^0\rightarrow l_i l_j$ are deduced here
\begin{eqnarray}
&&\Gamma(h^0\rightarrow l_il_j)=(h^0\rightarrow \bar{l}_il_j)+(h^0\rightarrow l_i\bar{l}_j),
\end{eqnarray}
where $\Gamma\left(h^0\rightarrow \bar{l}_il_j\right)=\frac{1}{16\pi}m_{h^0}\left(|N_L|^2+|N_R|^2\right)$\cite{h01,h02}. Correspondingly, the calculations for $h^0\rightarrow l_i\bar{l}_j$ are same as those for $h^0\rightarrow \bar{l}_il_j$.

Above all, the branching ratios of $h^0\rightarrow l_il_j$ can be summarized as
\begin{eqnarray}
&&Br(h^0\rightarrow l_il_j)=\Gamma(h^0\rightarrow l_il_j)/\Gamma_{h^0}.
\end{eqnarray}
Here, the total decay width of the 125.1 GeV Higgs boson is $\Gamma_{h^0}\simeq4.1\times10^{-3}$ GeV\cite{PDG}.

$B^0$ meson is made up of $d$ $\bar{b}$ and $B_s^0$ meson is constituted of $s$ $\bar{b}$. The present experiment upper bounds for $B^0$ and $ B_s^0$ meson decays are respectively $Br(B^0\rightarrow e^+\mu^-)<2.8\times10^{-9}$ and $Br(B_s^0\rightarrow e^+\mu^-)<1.1\times10^{-8}$\cite{PDG}. New contributions to rare $B^0$ and $ B_s^0$ meson decays emerge at one-loop level with the box diagrams. In the EBLMSSM, the redefined particles sleptons and sneutrinos lead to new effects to these rare $B^0$ and $ B_s^0$ meson decays. So parameters $\tan\beta_{NL}$ and $v_{Nlt}$ may play the dominated roles to the $B^0$ and $B_s^0$ meson decays.

$\pi^+,K^+$ mesons are respectively comprised of $u$ $\bar{d}$ and $u$ $\bar{s}$. Particle Date Group gives us the present experiment upper bounds for $(\pi^+/K^+)\rightarrow l_i^+\nu_j$, which are $Br(\pi^+\rightarrow \mu^+\nu_e)<8.0\times10^{-3}$ and $Br(K^+\rightarrow \mu^+\nu_e)<4.0\times10^{-3}$\cite{PDG}. In the EBLMSSM, the penguin type diagrams, self-energy type diagrams and box type diagrams all affect the processes $(\pi^+/K^+)\rightarrow l_i^+\nu_j, i\neq j$. CLFV contributions arise from loop corrections with the $W^{\pm}$ and heavy charged Higgs propagator. Furthermore, the loop contributions are also related with the exotic slepton (sneutrino), exotic lepton (neutrino), lepton neutralino and slepton (sneutrino) particles. Therefore, processes $(\pi^+/K^+)\rightarrow l_i^+\nu_j$ will be strongly affected by parameters presented in the EBLMSSM. We hope a detailed analysis is going to be discussed in our next work.
\section{Numerical Results}
In this section, we discuss the numerical results. In our previous work\cite{EBL}, we research the processes $h^0\rightarrow \gamma\gamma$, $h^0\rightarrow VV, V=(Z,W)$ in the EBLMSSM, and the corresponding numerical results are discussed in section 5.1 of work\cite{EBL}. The CP-even Higgs masses $m_{h^0}, m_{H^0}$ and CP-odd Higgs mass $m_A^0$ are also analyzed. In the reasonable parameter space, the values of branching ratios for $h^0\rightarrow \gamma\gamma$($R_{\gamma\gamma}$) and $h^0\rightarrow VV$($R_{VV}$) both meet the experiment limits. Therefore, the Higgs decays in the EBLMSSM play important roles to promote physicists to explore new physics. And the corresponding constraints are also considered in our work. The CP-even Higgs mass is considered as an input parameter, which is $m_{h^0}=125.1$ GeV in our latter numerical discussions.

In the EBLMSSM, to obtain a more transparent numerical results, we adopt the following assumptions on parameter space:
\begin{eqnarray}
&&Y_{u_4} = 1.2Y_t,~Y_{u_5} = 0.6Y_t,~
Y_{d_4}=Y_{d_5} = 2Y_b,~Y_{\nu_4} = Y_{\nu_5} = 0.8,~\mu_B = \mu_L = 0.5 {\rm TeV},\nonumber\\&&
m_{\tilde{Q}_4} = m_{\tilde{Q}_5} =m_{\tilde{U}_4} = m_{\tilde{U}_5} = m_{\tilde{D}_4} = m_{\tilde{D}_5} = m_{\tilde{\nu}_4} = m_{\tilde{\nu}_5} =1{\rm TeV},~B_4 = L_4 = 1.5,\nonumber\\&&
A_{u_4} =A_{u_5} = A_{d_4} = A_{d_5} =A_{\nu_4} = A_{\nu_5} =1{\rm TeV},
\lambda_Q = 0.4,\lambda_u=\lambda_d = 0.5,(\lambda_{Nc})_{ii} = 1,\nonumber\\&&
A_{BQ}=A_{BU} = A_{BD} = 1{\rm TeV},~g_B = 1/3,~g_L = 1/6,~\tan\beta_B =1.5,~\tan\beta_L  = 2,\nonumber\\&&
m_{\tilde{Q}_3}=m_{\tilde{U}_3}= 1.2{\rm TeV},~m_{\tilde{D}_3}= 1.5{\rm TeV},~A_t = 1.7{\rm TeV},~A_b = 3{\rm TeV},~M_L = 1{\rm TeV},
\nonumber\\&&(m_{\tilde{\nu}})_{ii}= 1{\rm TeV},~~(A_{N})_{ii} =(A_{Nc})_{ii} = 0.5{\rm TeV},~~m_{\Lambda}=1{\rm TeV} \label{canshu}
\end{eqnarray}
where $i=1,2,3$, $Y_t$ ($Y_b$) corresponds to the Yukawa coupling constant of top (bottom) quark, whose concrete form can be written as
$Y_t = \sqrt{2} m_t/(\upsilon \sin\beta)$ ($Y_b = \sqrt{2} m_b/(\upsilon \cos\beta)$).

In order to simplify the numerical analysis, we use the following assumptions:
\begin{eqnarray}
&&m_{\tilde{L}_4} = m_{\tilde{L}_5} =m_{\tilde{e}_4} = m_{\tilde{e}_5} = M_{\tilde{E}},A_{e_4} =A_{e_5} = A_{\tilde{E}},\lambda_L =\lambda_E=\lambda_{N_L} = L_l,\nonumber\\&&A_{LL}=A_{LE} = A_{LN} = A_E,(\lambda_4^2)^{IJ} =(\lambda_6^2)^{IJ} = (Lm^2)^{IJ},I,J=1,2,3,v_{Nlt}=v_N,\nonumber\\&&
(m_{\tilde{L}}^2)_{ii}=(M_{Ls})_{ii}^2,~(m_{\tilde{L}}^2)_{ij}=M_{Lf}^2,~(A_l)_{ii}=Al,(A'_l)_{ii}=A'l,i,j=1.2.3,i\neq j.
\end{eqnarray}
We take $\sqrt{(Lm^2)^{12}}=L_F$ and $\sqrt{(Lm^2)^{13}}=\sqrt{(Lm^2)^{23}}=L_f$.
\subsection{$l_j\rightarrow l_i \gamma$}
\subsubsection{$\mu\rightarrow e \gamma$}
CLFV process $\mu\rightarrow e \gamma$ contributes to explore the new physics, whose experiment upper bound of the branching ratio is around $5.7 \times10^{-13}$ at $90\%$ confidence level. In this part, we discuss the effects on process $\mu\rightarrow e \gamma$ from some new introduced parameters in the EBLMSSM.

\begin{figure}[t]
\centering
\includegraphics[width=7cm]{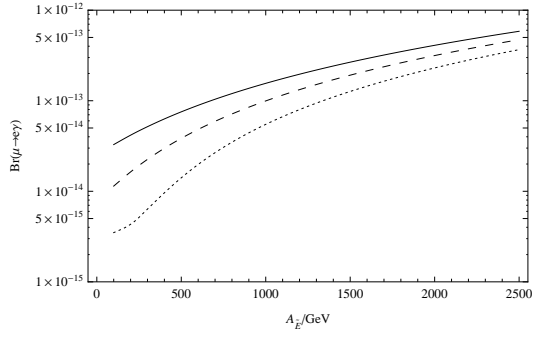}\\
\caption{ With $Y_{e5}=1.0(1.5,2.0)$, the branching ratios of $\mu\rightarrow e \gamma$ versus parameter $A_{\tilde {E}}$ are plotted by the dotted, dashed and solid lines respectively.} \label{fig4}
\end{figure}

Parameter $A_{\tilde{E}}$ is present in the non-diagonal parts of the exotic slepton mass matrix. Parameter $Y_{e5}$ is not only related to the non-diagonal parts of the exotic lepton and exotic slepton mass matrices, but also connected with the diagonal exotic slepton elements. In the EBLMSSM, exotic lepton and exotic slepton are both different from those in the BLMSSM. We assume that $M_{\tilde{E}}=\mu_Y=1.5$TeV, $A_E=\mu_{NL}=1$TeV, $L_l=1$, $L_F=10^{-3}$, $\mu=0.7$TeV, $B_Y=0.94$TeV, $Y_{e4}=0.5$, $(M_{Ls})_{11}^2=6 {\rm TeV}^2$, $(M_{Ls})_{22}^2=4 {\rm TeV}^2$, $(M_{Ls})_{33}^2=1 {\rm TeV}^2$, $M_{Lf}^2=10^{-3} {\rm TeV}^2$, $Al=2$TeV, $A'l=0.3$TeV, $m_1=m_2=1.5{\rm TeV}$, $\tan\beta=6$ and $\tan\beta_{NL}=2$ . With $Y_{e5}=1.0(1.5,2.0)$, the branching ratios of $\mu\rightarrow e \gamma$ versus parameter $A_{\tilde {E}}$ are studied, which are shown in FIG.\ref{fig4}.
When $A_{\tilde{E}}$ is in the region $0.1 \sim 2.5$TeV, the numerical results change from $5\times10^{-15}$ to $5\times10^{-13}$. These three lines all increase quickly and approach to the experiment upper bound. Therefore, $A_{\tilde{E}}$ affects the numerical results strongly. Furthermore, corresponding to same $A_{\tilde {E}}$, the solid line results are about 2 times as the dashed line results, and the dashed line results are about 2 times as the dotted line results. Larger $Y_{e5}$ can lead to larger numerical results.
\begin{figure}[t]
\centering
\includegraphics[width=7cm]{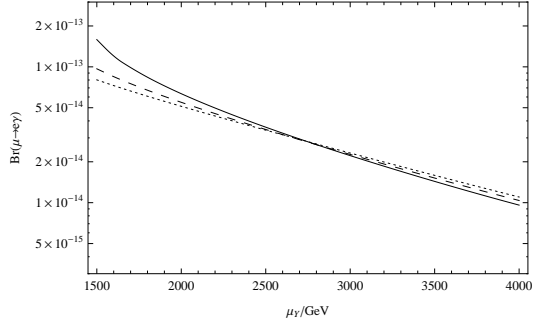}\\
\caption{ With $B_Y=0.4(0.8,1.2)$ TeV, the branching ratios of $\mu\rightarrow e \gamma$ versus parameter $\mu_Y$ are plotted by the dotted, dashed and solid lines respectively.} \label{fig5}
\end{figure}

With the introduced superfields $Y$ and $Y'$ in the EBLMSSM, we deduce the $Y$ and $\tilde{Y}$ mass matrices. Parameters $\mu_Y$ and $B_Y$ are respectively present in the diagonal and non-diagonal terms of the $Y$ mass matrix. And the mass of $\tilde{Y}$ possesses the same value as $\mu_Y$. So these two parameters affect the $Y$-lepton-exotic lepton and $\tilde{Y}$-lepton-exotic slepton couplings. Furthermore, these new couplings make contributions to the numerical results. Using $Y_{e4}=0.8$, $Y_{e5}=1.5$ and $A_{\tilde{E}}=1$TeV, we plot the branching ratios changing with $\mu_Y$ in FIG.\ref{fig5}. The dotted (dashed, solid) line represents $B_Y=0.4(0.8,1.2)$TeV. We find that the branching ratios decrease quickly with the increasing $\mu_Y$, which indicates that the large $\mu_Y$ can restrain the numerical results evidently. Furthermore, the numerical results of these three lines are almost same with the unchanging $\mu_Y$, so the contributions from parameter $B_Y$ is small.
\begin{figure}[t]
\centering
\includegraphics[width=7cm]{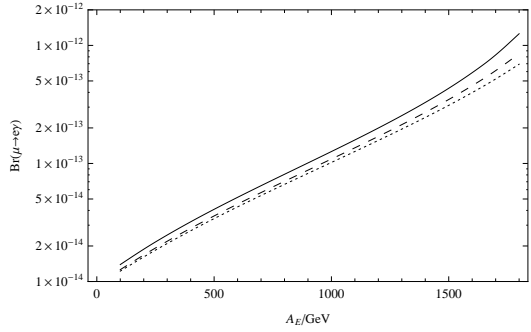}\\
\caption{ With $\mu_{NL}=0.7(1.0,1.3)$ TeV, the branching ratios of $\mu\rightarrow e \gamma$ versus parameter $A_E$ are plotted by the dotted, dashed and solid lines respectively.} \label{fig6}
\end{figure}

Then, we study effects from the parameters $A_E$ and $\mu_{NL}$ on our numerical results. In EBLMSSM, parameters $A_E$ and $\mu_{NL}$ are both the non-diagonal elements in the exotic slepton and exotic sneutrino mass matrices. $\mu_{NL}$ is also the non-diagonal element of lepton neutralino mass matrix. In FIG.\ref{fig6}, we present the branching ratios of $\mu\rightarrow e \gamma$ versus $A_E$ with $\mu_{NL}=0.7(1.0,1.3)$TeV, and the concrete results are plotted by dotted (dashed, solid) line. These three lines all increase quickly when $A_E$ ranges from 0.1 to 1.8 TeV. Therefore, as the sensitive parameters in the EBLMSSM, the large $A_E$ produces the large contributions on the results. However, the numerical results slightly decrease with the enlarging $\mu_{NL}$, and the effects from $\mu_{NL}$ are not so obvious as that $A_E$.
\subsubsection{$\tau\rightarrow \mu \gamma$ ($\tau\rightarrow e\gamma$)}
In a similar way, the CLFV processes $\tau\rightarrow e \gamma$ and $\tau\rightarrow \mu \gamma$ are studied. The corresponding experimental upper bounds of the branching ratios are $Br(\tau\rightarrow e \gamma)<3.3\times 10^{-8}$ and $Br(\tau\rightarrow \mu \gamma)<4.4\times 10^{-8}$.
\begin{figure}[t]
\centering
\includegraphics[width=8cm]{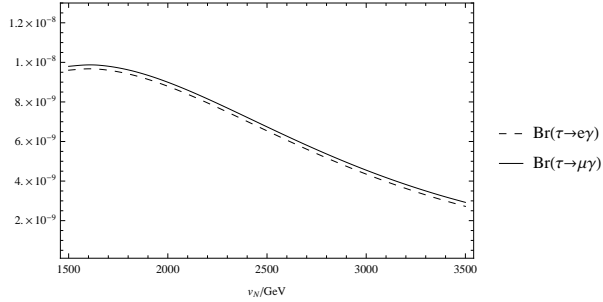}\\
\caption{The branching ratios of $\tau\rightarrow e \gamma$ and $\tau\rightarrow \mu \gamma$ versus parameter $v_{N}$ are plotted by the dashed line and solid line respectively.} \label{fig7}
\end{figure}

As a new introduced parameter in the EBLMSSM, parameter $v_{N}$ is present in the mass matrices of slepton, sneutrino, exotic lepton (neutrino), exotic slepton (sneutrino) and lepton neutralino. In this part, we research the branching ratios of  $\tau\rightarrow e \gamma$ and $\tau\rightarrow \mu \gamma$ changing with $v_{N}$. Supposing $M_{\tilde{E}}=\mu_Y=1.5$TeV, $A_E=\mu_{NL}=1$TeV, $A_{\tilde{E}}=1$TeV, $L_l=1$, $L_f=0.08$, $\mu=0.7$TeV, $B_Y=0.94$TeV, $Y_{e4}=Y_{e5}=0.8$, $(M_{Ls})_{ii}^2=S_m^2=1 {\rm TeV}^2$, i=1,2,3, $M_{Lf}^2=10^{-2} {\rm TeV}^2$, $Al=2$TeV, $A'l=0.3$TeV, $m_1=m_2=1.5{\rm TeV}$, $\tan\beta=6$ and $\tan\beta_{NL}=2$, we plot the numerical results with $v_{N}$ in FIG.\ref{fig7} by dashed line and solid line respectively. Obviously, when the values of $v_{N}$ change from 1.5 to 3.5 TeV, the results of $Br(\tau\rightarrow e \gamma)$ and $Br(\tau\rightarrow \mu \gamma)$ both shrink quickly. This implies that $v_{N}$ is a sensitive parameter. Though the figure of process $\tau\rightarrow e \gamma$ is under that of $\tau\rightarrow \mu \gamma$, the both lines possess almost the same results when $v_{N}$ takes same value. So we only study the branching ratios of process $\tau\rightarrow \mu \gamma$ in following discussion.
\begin{figure}[t]
\centering
\includegraphics[width=7cm]{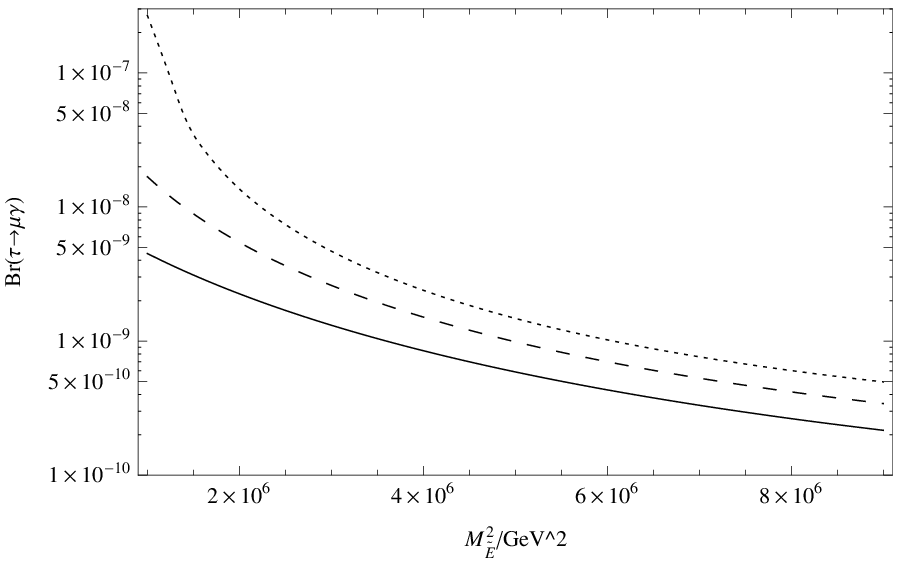}\\
\caption{With $L_l=0.7(1.0,1.3)$, the branching ratios of $\tau\rightarrow \mu \gamma$ versus parameter $M_{\tilde{E}}^2$ are plotted by the dotted, dashed and solid lines respectively.} \label{fig8}
\end{figure}
\begin{figure}[t]
\centering
\includegraphics[width=7cm]{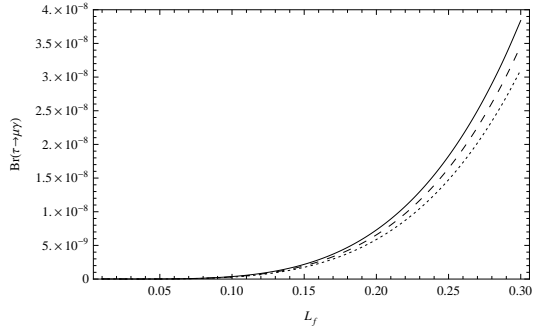}\\
\caption{With $Y_{e4}=0.5(1.0,1.5)$, the branching ratios of $\tau\rightarrow \mu \gamma$ versus parameter $L_f$ are plotted by the dotted, dashed and solid lines respectively.} \label{fig9}
\end{figure}

Appearing in the diagonal terms of the exotic slepton and exotic sneutrino mass squared matrices, $M_{\tilde{E}}$ affects the $\tilde{Y}$-lepton-exotic slepton coupling. Parameter $L_l$, not only in the exotic lepton (neutrino) but also in exotic slepton (sneutrino) mass matrices, produces contributions to the numerical results through $Y$-lepton-exotic lepton and $\tilde{Y}$-lepton-exotic slepton couplings. With $(M_{Ls})_{11}^2=6 {\rm TeV}^2$, $(M_{Ls})_{22}^2=4 {\rm TeV}^2$, $(M_{Ls})_{33}^2=1 {\rm TeV}^2$ and $v_{N}=3$TeV, the numerical results versus $M_{\tilde{E}}^2$ are plotted in FIG.\ref{fig8}. The dotted (dashed, solid) line corresponds to $L_l=0.7(1.0,1.3)$. The figure shows that these three lines all decrease quickly when $M_{\tilde{E}}^2$ varies from $1\times10^{6}$ to $9\times10^{6}{\rm GeV}^2$. With the same $M_{\tilde{E}}^2$, the branching ratio decreases remarkably when $L_l$ increases. Especially, the line is much steeper with $L_l=0.7$ than that $L_l=1.0(1.3)$. Obviously, both $M_{\tilde{E}}$ and $L_l$ are sensitive parameters to our numerical results.

Parameter $L_f$ influences the numerical results through $Y$-lepton-exotic lepton and $\tilde{Y}$-slepton-exotic slepton couplings. And parameter $Y_{e4}$ affects our numerical results through exotic lepton and exotic slepton. We discuss the numerical results with $L_f$ varying from 0.01 to 0.3 in FIG.\ref{fig9}. The dotted (dashed, solid) line corresponds to $Y_{e4}=0.5(1.0,1.5)$. The branching ratios possess slight changes when $Y_{e4}$ takes different values for the unchanged $L_f$, which indicates the effects from $Y_{e4}$ can be ignored in our following discussion. It is easy to see that the numerical results increase sharply with the enlarging $L_f$. So the non-diagonal elements of parameters $\lambda_4^2$ and $\lambda_6^2$ play important roles in our numerical studies.
\subsection{$\mu-e$ conversion rates in Au nuclei}
The present sensitivity for the muon conversion rates to electron in Au nuclei is $CR(\mu\rightarrow e:\;_{79}^{197}Au)<7\times 10^{-13}$. Considering the parameter constrains from $\mu\rightarrow e \gamma$, we analyze the numerical results for this $\mu-e$ conversion in Au nuclei.
\begin{figure}[t]
\centering
\includegraphics[width=7cm]{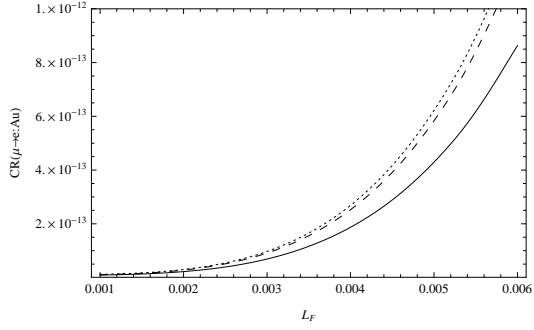}\\
\caption{With $\tan\beta_{NL}=1.8,2.2,2.6$, the $\mu-e$ conversion rates in Au nuclei versus parameter $L_F$ are plotted by the dotted, dashed and solid lines respectively.} \label{CueLF}
\end{figure}
\begin{figure}[t]
\centering
\includegraphics[width=7cm]{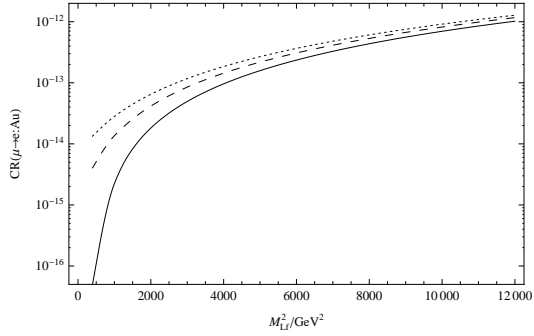}\\
\caption{With $\mu=0.4(0.5,0.6)$TeV, the $\mu-e$ conversion rates in Au nuclei versus parameter $M_{Lf}^2$ are plotted by the dotted, dashed and solid lines respectively.} \label{CueMLf}
\end{figure}

As the non-diagonal elements of matrix $(Lm^2)^{IJ}$ in the EBLMSSM, $\sqrt{(Lm^2)^{12}}=L_F$ affects the numerical results through exotic lepton and exotic slepton. Choosing  $M_{\tilde{E}}=A_E=\mu_{NL}=\mu=1$TeV, $\mu_Y=2$TeV, $L_l=0.8$, $B_Y=1.5$TeV, $Y_{e4}=0.8$, $Y_{e5}=1.5$, $S_m^2=6 {\rm TeV}^2$, $M_{Lf}^2=500 {\rm GeV}^2$, $m_1=m_2=3{\rm TeV}$, $Al=2$TeV, $A'l=0.3$TeV and $\tan\beta=6$, we analyze the $\mu-e$ conversion rates in Au nuclei with $L_F$ in FIG.\ref{CueLF}. $\tan\beta_{NL}=1.8,2.2,2.6$ correspond to the dotted, dashed and solid lines respectively. When $L_F$ changes from 0.001 to 0.006, These three lines all enlarge quickly and can easily reach the present sensitivity. So $L_F$ greatly contributes to the numerical results. Furthermore, as $L_F$ takes the same value, the larger $\tan\beta_{NL}$, the smaller numerical result it is. The above analyses indicate $L_F$ and $\tan\beta_{NL}$ are both sensitive parameters.

As the non-diagonal elements of slepton and sneutrino mass matrices, $M_{Lf}$ lead to strong mixing for slepton (sneutrino) with different generations. The parameter $\mu$ presents in the mass matrices of slepton, sneutrino, exotic slepton and exotic sneutrino. So we study the $\mu-e$ conversion rates in Au nuclei versus parameters $M_{L_f}^2$. As $M_{\tilde{E}}=2$TeV, $\mu_Y=2$TeV, $B_Y=0.9$TeV, $S_m^2=12 {\rm TeV}^2$, $m_1=m_2=3{\rm TeV}$, $Al=1.9$TeV, $\tan\beta_{NL}=2$ and $L_F=0.001$, we show the numerical results changing with $M_{Lf}^2$, which are given in FIG.\ref{CueMLf}. The dotted, dashed and solid lines respectively correspond to $\mu=0.4,0.5,0.6$TeV. With the enlarging $M_{Lf}^2$, the numerical results increase quickly. As $M_{Lf}^2>5000{\rm GeV}^2$ and taking the same values, these three lines almost possess similar results, which indicates that the effects from parameter $\mu$ are small.
\subsection{$\tau$ decays}
The experiment upper bounds for $\tau$ decays are $Br(\tau\rightarrow 3e)<2.7\times 10^{-8}$ and $Br(\tau\rightarrow 3\mu)<2.1\times 10^{-8}$. Considering the constraints from $\tau\rightarrow e \gamma$ and $\tau\rightarrow \mu \gamma$, we discuss the numerical results for decays $\tau\rightarrow 3e$ and $\tau\rightarrow 3\mu$.

\begin{figure}[t]
\centering
\includegraphics[width=8cm]{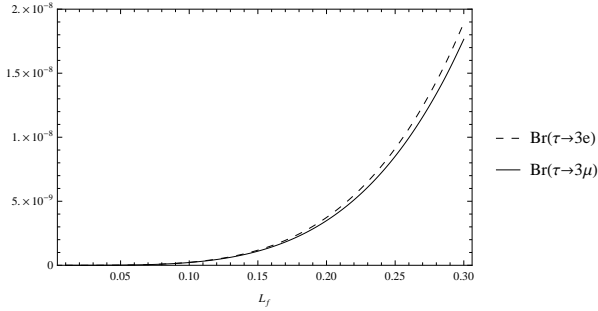}\\
\caption{The branching ratios of $\tau\rightarrow 3\mu$ and $\tau\rightarrow 3e$ versus parameter $L_f$ are plotted by the solid line and dashed line respectively.} \label{tauLf}
\end{figure}

Using $\mu_Y=3$TeV, $M_{\tilde{E}}=A_E=\mu_{NL}=1$TeV, $L_l=0.8$, $\mu=0.7$TeV, $B_Y=1.5$TeV, $Y_{e4}=0.8$, $Y_{e5}=1.5$, $(M_{Ls})_{ii}^2=6 {\rm TeV}^2$, $M_{Lf}^2=500 {\rm GeV}^2$, $Al=1$TeV, $A'l=0.3$TeV, $m_1=m_2=0.5{\rm TeV}$, $\tan\beta=6$ and $\tan\beta_{NL}=1.5$, we plot the numerical results of $\tau\rightarrow 3e$ and $\tau\rightarrow 3\mu$ in FIG.\ref{tauLf} by dotted line and solid line respectively. With parameter $L_f$ changing from 0.01 to 0.3, the values of these two lines are almost the same and both increase quickly. So we just discuss the numerical results for $\tau\rightarrow 3e$ decays as follows.
\begin{figure}[t]
\centering
\includegraphics[width=7cm]{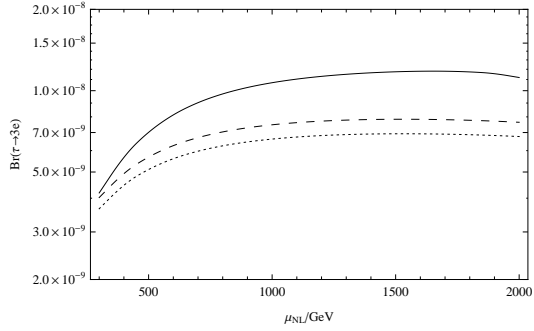}\\
\caption{With $\tan\beta_{NL}=1.5(1.7,1.9)$, the branching ratios of $\tau\rightarrow 3e$ versus parameter $\mu_{NL}$ are plotted by the dotted, dashed and solid lines respectively.} \label{muNLtbNL}
\end{figure}

Choosing $Y_{e4}=0.8$, $Y_{e5}=1.5$, $M_{\tilde{E}}=\mu_Y=1.5$TeV and $B_Y=0.9$TeV, we study the branching ratios of $\tau\rightarrow 3e$ changing with parameter $\mu_{NL}$. The numerical results varying with $\tan\beta_{NL}=1.5(1.7,1.9)$ are plotted by the dotted, dashed and solid lines respectively. As $\mu_{NL}$ changes from 0.3 TeV to 1 TeV, these three lines all have the obviously improvement. As $\mu_{NL}>1$ TeV, the numerical results increase slowly. Besides, when $\mu_{NL}$ dose not change, the branching ratios of $\tau\rightarrow 3e$ enlarge with the increased $\tan\beta_{NL}$, and the bigger $\tan\beta_{NL}$, the bigger change it is in the graph. Therefore, both $\mu_{NL}$ and $\tan\beta_{NL}$ affect the numerical results in a certain degree.
\subsection{$h^0\rightarrow l_i l_j$}
In this part, we study the CLFV processes $h^0\rightarrow l_i l_j$. The most strict constraint $m_{h^0}=125.1$GeV is considered as an input parameter. We also take into account the limits from processes $l_j\rightarrow l_i \gamma$, the muon conversion to electron in Au nuclei and the $\tau$ decays discussed above.
\subsubsection{$h^0\rightarrow \mu \tau$$(h^0\rightarrow e \tau)$}
\begin{figure}[t]
\centering
\includegraphics[width=8cm]{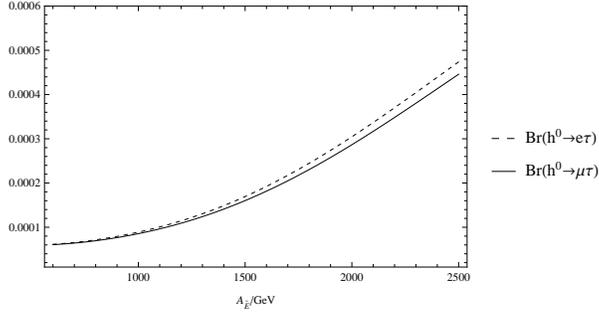}\\
\caption{The branching ratios of $h^0\rightarrow \mu \tau$ and $h^0\rightarrow e \tau$ versus parameter $A_{\tilde{E}}$ are plotted by the solid line and dashed line respectively.} \label{fig10}
\end{figure}

At first, we picture the branching ratios of decays $h^0\rightarrow \mu \tau$ and $h^0\rightarrow e \tau$ versus $A_{\tilde{E}}$ in FIG.\ref{fig10}. We choose the relevant parameters as $\mu_Y=1.5$TeV, $M_{\tilde{E}}=1.4$TeV, $A_E=m_1=1$TeV, $\mu_{NL}=2$TeV, $L_l=1$, $(Lm^2)^{13}=(Lm^2)^{23}=L_f=0.3^2$, $\mu=0.7$TeV, $B_Y=0.94$TeV, $Y_{e4}=0.8$, $Y_{e5}=1.5$, $S_m^2=1 {\rm TeV}^2$, $M_{Lf}^2=12000 {\rm GeV}^2$, $Al=1$TeV, $A'l=3$TeV, $m_2=0.5$TeV, $\tan\beta=6$ and $\tan\beta_{NL}=2$. Although the line of $h^0\rightarrow \mu \tau$ is under that of $h^0\rightarrow e \tau$, these two processes almost have the same variation trend. With the enlarging $A_{\tilde{E}}$, the numerical results increase quickly.
\begin{figure}[t]
\centering
\includegraphics[width=7cm]{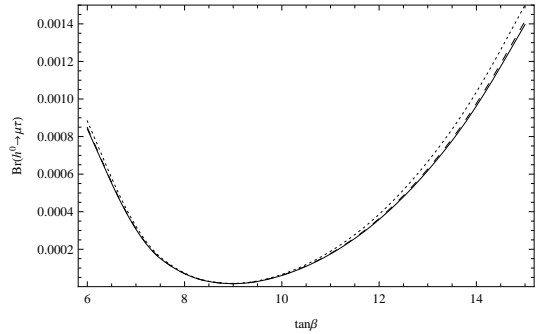}\\
\caption{With $S_m^2=1(2,3){\rm TeV}^2$, the branching ratios of $h^0\rightarrow \mu \tau$ versus parameter $\tan{\beta}$ are plotted by the dotted, dashed and solid lines respectively.} \label{fig12}
\end{figure}

Then the effects from the parameters $\tan{\beta}$ and $S_m$ are studied.
$\tan{\beta}$ is related to $v_u$ and $v_d$, and appears in almost all mass matrices of CLFV processes. $S_m$ are present in the diagonal elements of slepton and sneutrino mass matrices. With $A_{\tilde{E}}=2$TeV, $L_f=0.25$, $Y_{e4}=1.2$ and $Y_{e5}=0.8$, FIG.\ref{fig12} shows the branching fractions of $h^0\rightarrow \mu \tau$ varying with the parameter $\tan{\beta}$. $S_m^2=1(2,3){\rm TeV}^2$ corresponds to the dotted (dashed, solid) line. These three lines almost overlap, so the effects from $S_m$ are small. As $\tan{\beta}$ varies from 6 to 9, the numerical results decrease obviously. As $\tan{\beta}>9$, the numerical results increase quickly. So $\tan{\beta}$ plays very important roles to CLFV processes.
\subsubsection{$h^0\rightarrow e \mu$}
\begin{figure}[t]
\centering
\includegraphics[width=8cm]{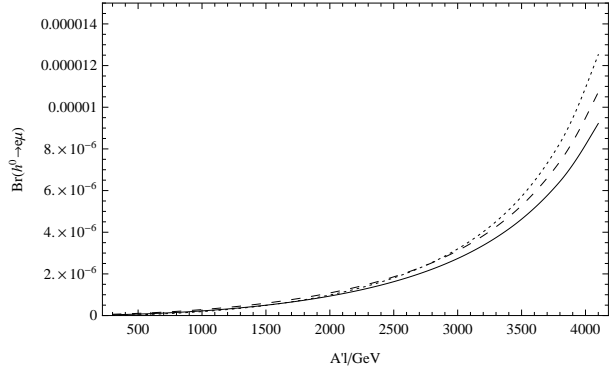}\\
\caption{ With $Al=0.5(1,1.5)$TeV, the branching ratios of $h^0\rightarrow e \mu$ versus parameter $A'l$ are plotted by the dotted (dashed, solid) line.} \label{fig13}
\end{figure}

The latest experiment upper bound of decay $h^0\rightarrow e \mu$ is smaller than $0.035\%$ at $95\%$ confidence level, which is detected by the CMS Collaboration. $Al$ and $A'l$ both appear in the non-diagonal terms of the slepton mass matrix. Considering the constraints from $\mu\rightarrow e \gamma$ and $\mu-e$ conversion in Au nuclei, we take $M_{\tilde{E}}=A_E=\mu_{NL}=1$TeV, $A_{\tilde{E}}=\mu_Y=1.5$TeV $L_l=1$, $L_F=0.006$, $B_Y=0.94$TeV, $Y_{e4}=1.5$, $Y_{e5}=0.8$, $S_m^2=1 {\rm TeV}^2$, $M_{Lf}^2=12000 {\rm GeV}^2$, $m_1=m_2=0.5$TeV, $\mu=0.7$TeV, $\tan\beta=6$ and $\tan\beta_{NL}=2$. The dotted (dashed, solid) line in FIG.\ref{fig13} denotes the branching ratios of $h^0\rightarrow e \mu$ versus $A'l$ with $Al=0.5(1,1.5)$TeV. These three lines all increase quickly with the enlarging $A'l$. So $A'l$ play important roles to the numerical results. Although the larger $Al$, the smaller numerical results they are, the contributions from $Al$ are very weak.
\begin{figure}[t]
\centering
\includegraphics[width=8cm]{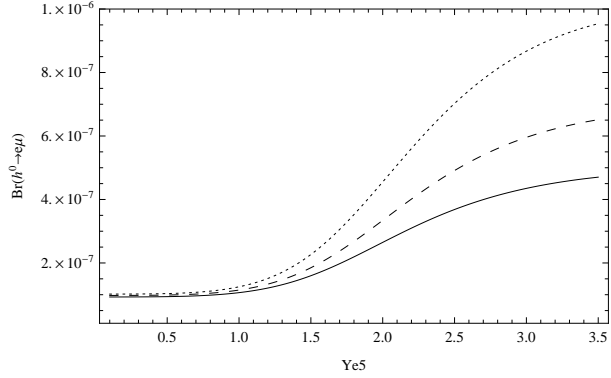}\\
\caption{ With $M_{\tilde{E}}=1.1(1.4,1.7)$TeV, the branching ratios of $h^0\rightarrow e \mu$ versus parameter $Y_{e5}$ are plotted by the dotted, dashed and solid lines respectively.} \label{fig15}
\end{figure}

At last, we discuss the effects from parameters $Y_{e5}$ and $M_{\tilde{E}}$. With $m_1=m_2=0.5$TeV, $Al=1.5$TeV, $Y_{e4}=0.5$ and $L_F=0.006$, the branching ratios varying with $Y_{e5}$ are ploted in FIG.\ref{fig15}. The dotted, dashed and solid lines respectively correspond to $M_{\tilde{E}}=1.1,1.4,1.7$TeV. These three lines all slightly increase when $Y_{e5}$ varies from 0.1 to 1.0. As $Y_{e5}$ still increases from 1.0, the results have much more conspicuous enlargement. However, the total contributions from $Y_{e5}$ are not so obvious. With the enlarging $M_{\tilde{E}}$, the numerical results reduce more and more slowly.
\section{discussion and conclusion}
We add exotic superfields $\Phi_{NL}$, $\varphi_{NL}$, $Y$ and $Y'$ to the BLMSSM,
and this new model is named as the EBLMSSM. In ${\cal W}_Y$, $\lambda_4(\lambda_6)$ is the coupling coefficient of $Y$-lepton-exotic lepton and $\tilde{Y}$-lepton-exotic slepton. We assume $\lambda_4^2=\lambda_6^2$ is a $3\times3$ squared matrix and its non-diagonal elements are related with the CLFV. Being different from the BLMSSM, the exotic slepton (sneutrino) of 4-th and 5-th generations mix together and form a $4\times4$ matrix. The Majorana particle, lepton neutralino $\chi_L^0$, is corrected to be a $5\times 5$ matrix due to the introduction of superpartners $\psi_{\Phi_{NL}}$ and $\psi_{\varphi_{NL}}$. The terms relating with exotic lepton (neutrino) and slepton (sneutrino) are also adjusted. In Section III, we show the corresponding mass matrices and couplings of the EBLMSSM. The EBLMSSM has more abundant contents than that BLMSSM for the lepton physics.

Considering the constraints from decays $h^0\rightarrow\gamma\gamma$ and $h^0\rightarrow VV,V=(Z,W)$, we study the CLFV processes $l_j\rightarrow l_i \gamma$, muon conversion to electron in Au nuclei and the $\tau$ decays in the framework of the EBLMSSM. Parameters $Y_{e5}$ and $M_{Lf}$ affect the numerical results in a certain degree. As the new introduced parameters in the EBLMSSM, $\mu_Y,\tan{\beta}_{NL}, A_{\tilde{E}},A_E,L_l,M_{\tilde{E}}$ and $v_N$ play important roles. Especially parameters $L_f$ and $L_F$ are all very sensitive parameters, which influence the numerical results very remarkably. FIG.\ref{fig9}, FIG.\ref{CueLF} and FIG.\ref{tauLf} indicate that the enlarging $L_f$ and $L_F$ can easily improve the numerical results. Then, the 125.1 GeV Higgs boson decays with CLFV $h^0\rightarrow l_il_j$ are discussed. As an important constraint, $m_{h^0}=125.1$GeV is regarded as an input parameter. Taking into account the constraints from the parameter space of decays $l_j\rightarrow l_i \gamma$, muon conversion to electron in Au nuclei and the $\tau$ decays, we analyze the numerical results for $h^0\rightarrow l_il_j$ in EBLMSSM. Parameters $\mu$ and $A'l$ affect the CLFV processes in a certain degree. The effects from $\tan\beta$ are very obvious. So $\tan\beta$ is a sensitive parameter. Above all, due to the new particles introduced in the EBLMSSM, the numerical results can easily approach to the present experiment upper bounds.

{\bf Acknowledgments}

This work is supported by the Major Project of National Natural Science Foundation of China (NNSFC) (No. 11535002, No. 11605037, No. 11647120, No. 11705045, No. 11275036),
the Natural Science Foundation of Hebei province with Grant
No. A2016201010 and No. A2016201069, and the Natural Science Fund of
Hebei University with Grants No. 2011JQ05 and No. 2012-
242, Hebei Key Lab of Optic-Electronic Information and
Materials, the midwest universities comprehensive strength
promotion project, the youth top-notch talent support program of the Hebei Province.
\section{Appendix}
In this section, we give out the corresponding one-loop integral functions, which are read as:
\begin{eqnarray}
&&I_1(x_1,x_2)=\frac{1}{16{\pi}^2}[-(\bigtriangleup+1+\ln{x_{\mu}})+\frac{x_2\ln{x_2}-x_1\ln{x_1}}{(x_2-x_1)}],
\nonumber\\&&I_2(x_1,x_2)=\frac{1}{32{\pi}^2}[\frac{3+2\ln{x_2}}{(x_2-x_1)}-\frac{2x_2+4x_2\ln{x_2}}{(x_2-x_1)^2}
+\frac{2x_2^2\ln{x_2}-2x_1^2\ln{x_1}}{(x_2-x_1)^3}],
\nonumber\\&&I_3(x_1,x_2)=\frac{1}{16{\pi}^2}[\frac{1+\ln{x_2}}{(x_2-x_1)}+\frac{x_1\ln{x_1}-x_2\ln{x_2}}{(x_2-x_1)^2}],
\nonumber\\&&I_4(x_1,x_2)=\frac{1}{96{\pi}^2}[\frac{11+6\ln{x_2}}{(x_2-x_1)}-\frac{15x_2+18x_2\ln{x_2}}{(x_2-x_1)^2}
+\frac{6x_2^2+18x_2^2\ln{x_2}}{(x_2-x_1)^3}\nonumber\\&&\hspace{2.2cm}+\frac{6x_1^3\ln{x_1}-6x_2^3\ln{x_2}}{(x_2-x_1)^4}],
\nonumber\\&&I_5(x_1,x_2)=\frac{1}{16{\pi}^2}[(\bigtriangleup+1+\ln{x_{\mu}})+\frac{x_2+2x_2\ln{x_2}}{(x_1-x_2)}
+\frac{x_2^2\ln{x_2}-x_1^2\ln{x_1}}{(x_2-x_1)^2}],
\nonumber\\&&G_1(x_1,x_2,x_3)=\frac{1}{16{\pi}^2}[\frac{x_1\ln{x_1}}{(x_1-x_2)(x_1-x_3)}+\frac{x_2\ln{x_2}}{(x_2-x_1)(x_2-x_3)}
+\frac{x_3\ln{x_3}}{(x_3-x_1)(x_3-x_2)}],
\nonumber\\&&G_2(x_1,x_2,x_3)=\frac{1}{16{\pi}^2}[-(\bigtriangleup+1+\ln{x_{\mu}})+\frac{x_1^2\ln{x_1}}{(x_1-x_2)(x_1-x_3)}
\nonumber\\&&\hspace{3.0cm}+\frac{x_2^2\ln{x_2}}{(x_2-x_1)(x_2-x_3)}+\frac{x_3^2\ln{x_3}}{(x_3-x_1)(x_3-x_2)}],
\end{eqnarray}
with $\Delta=\frac{1}{\epsilon}-r_{\epsilon}+\ln{4\pi}$.

 \end{document}